%% file: bullbo.tex
\documentstyle[prl,aps,preprint,epsf]{revtex}
\begin{document}
\draft
\input psfig_19

\title{Phase Transitions in a Nonequilibrium Percolation Model}
\author{Siegfried Clar$^1$, Barbara Drossel$^2$, Klaus Schenk$^1$, and Franz
  Schwabl$^1$}
\address{${}^1$Institut f\"ur Theoretische Physik, \\
  Physik-Department der Technischen Universit\"at M\"unchen, \\
  James-Franck-Str., D-85747 Garching, Germany}
\address{${}^2$ Department of Theoretical Physics, \\
  University of Manchester, \\
  Manchester M13 9PL, England} 
\date{\today} 
\maketitle
\begin{abstract}
  We investigate the percolation properties of a two--state (occupied -- empty)
  cellular automaton, where at each time step a cluster of occupied sites is
  removed and the same number of randomly chosen empty sites are occupied
  again.  We find a finite region of critical behavior, formation of
  synchronized stripes, additional phase transitions, as well as violation of
  the usual finite--size scaling and hyperscaling relations, phenomena that are
  very different from conventional percolation systems.  We explain the
  mechanisms behind all these phenomena using computer simulations and analytic
  arguments.
\end{abstract}


\narrowtext

\section{Introduction}

During the past years, systems which exhibit self--organized criticality (SOC)
have attracted much attention, since they might explain part of the abundance
of fractal structures in nature \cite{bak87}. Their common features are slow
driving or energy input (e.g. dropping of sand grains \cite{bak87}, increase of
strain \cite{ola92}, tree growth \cite{dro92}, spontaneous mutations
\cite{sne93}) and rare dissipation events which are instantaneous on the time
scale of driving (e.g. sand avalanches, earthquakes, fires, or a series of
rapid mutations).  In the stationary state, the size distribution of
dissipation events obeys a power law, irrespective of initial conditions and
without the need to fine-tune parameters.  There is, however, no reason to
expect that systems with slow driving and instantaneous avalanches are always
SOC. Such systems might also have many small avalanches which release only
little energy, or only large avalanches which release a finite part of the
system's energy, or some combination of both. SOC systems are naturally at the
critical point, due to e.g. a conservation law (sandpile model), a second time
scale separation (forest--fire model), 
a competition between open boundary conditions and the tendency of 
neighboring sites to synchronize (earthquake model \cite{ola92,mid95} -- 
see however \cite{lis96} for a counterexample), 
or due to extremal dynamics ('evolution' model \cite{sne93}).  
Often, the critical behavior breaks down when details of the
model rules are changed (e.g. the boundary conditions in the earthquake model
\cite{mid95} or the tree growth rule in the forest--fire model \cite{dro96}).

There are certain parallels between these models and equilibrium critical
systems, since both consist of many small units which interact with their
neighbors, and since spin clusters in an Ising model or clusters of occupied
sites in percolation theory can be compared to avalanches. However, the
critical behavior of nonequilibrium systems can depend on microscopic details,
as mentioned above, in contrast to equilibrium critical phenomena, which
commonly show universal behavior. Also, nonequilibrium systems do not satisfy a
detailed-balance condition and can e.g.  show periodic behavior.  Furthermore,
avalanches are usually released when some variable reaches locally a threshold,
while other regions of the system might be far below the threshold, and
consequently not all parts of the system look equal. This can in particular
result in more than one diverging length scale, as in the earthquake model
\cite{mid95} or in the forest--fire model \cite{hon96}. By contrast, in
equilibrium systems the energy is an extensive variable, which means that all
regions (which are large compared to the lattice constant) are equal. This is
e.g.  the basis for hyperscaling relations.

In this paper, we discuss in detail a model \cite{cla95} that belongs to the
mentioned class of nonequilibrium systems with avalanche--like dynamics. It is
a nonequilibrium percolation model, where clusters of occupied sites are
removed, and the same number of sites that have become empty are occupied at
random. The density of occupied sites is the control parameter of the model.
The 'avalanches' of our model are removal events, and the size of an
avalanche is the size of a removed cluster.  This model illustrates well the
fundamental differences between equilibrium and nonequilibrium, showing various
features that are not observed in equilibrium systems: The region of small
avalanches and the region of infinite avalanches are separated by a finite
region of critical behavior, where the correlation length diverges slower than
the system size. The exponent that relates the system size with the correlation
length depends on the density.  Besides the correlation length, there are other
relevant length scales.  Since the critical behavior occurs over a finite
density interval, the system can exhibit power laws naturally, without fine
tuning of parameters to a precise value. Therefore, our model belongs to the
class of SOC systems.  In the region of infinite avalanches, the system shows
synchronization with a period that depends on the value of the density. We
illustrate and explain all these observations using computer simulations and
analytical arguments. Part of the results were already published in
\cite{cla95}.

The work is structured as follows: In Sec.~\ref{definition}, we define the
model.  In Sec.~\ref{subcritical}, the subcritical phase and the critical point
of the model are treated. The mechanism that leads to criticality and the value
of the critical density are explained, and the exponent of the cluster size
distribution in 1D is calculated analytically.  Sec.~\ref{criticalregion}
discusses the critical phase.  The reason for the existence of a whole critical
phase as well as for its properties like non-standard finite-size scaling and
violation of hyperscaling are explained. The supercritical phase is treated in
Sec.~\ref{supercritical}.  First, we explain the existence of synchronized
stripes and their relation to the subcritical phase (Subs.~\ref{synchro}), then
we discuss hysteresis and the maximum possible number of stripes
(Subs.~\ref{hysteresis}), and, finally, we investigate stability, movement and
roughness of stripes (Subs.~\ref{roughness}).  In the conclusion
Sec.~\ref{conclusion}, we summarize and discuss our work.

\section{Definition of the Model}
\label{definition}

The non--equilibrium percolation model model is defined on a $d$--dimensional
hyper--cubic lattice with $L^d$ sites. Each site is either occupied or empty.
The control parameter of the system is the density of occupied sites, $\rho$.

The dynamics are defined by the following rules: (i) An occupied site in the
system is chosen at random and the whole cluster of $s$ occupied sites
connected to this site (by nearest-neighbor coupling) is removed from the
system, i.e. the occupied sites of that cluster turn to empty sites. (ii) We
occupy $s$ randomly chosen empty sites (possibly also including sites which
have become empty due to the removal of the cluster). (iii) Proceed with (i).
 
These rules ensure that the density of occupied sites $\rho$ is a conserved
quantity. Starting with a random initial state, the system approaches after a
transient time a stationary state that is characterized by a certain size
distribution of clusters, and where time average and ensemble average of all 
quantities are identical. Throughout this paper, we discuss only the properties
of the stationary state. These properties, that are explained in detail in the
next three sections, are as follows: For small densities, there are only small
clusters of occupied sites in the system. With increasing density, the size of
the largest cluster increases, and it diverges at a critical density $\rho_c$.
For $\rho_c < \rho < \rho_c^{(2)}$, the system is critical, i.e. the cluster
size distribution is a power law. The size of the largest cluster diverges
slower than the system size.  For $\rho > \rho_c^{(2)}$, the system has a
finite number of regions of different density. The region with the highest
density has a spanning cluster.

One can think of the dynamics of this model as 'explosions' that take place
at a randomly chosen site. During the explosion, the whole cluster connected to
the explosion site is blown up, and its constituents settle down somewhere else
in the system. Or one might think of colonies of animals which are dispersed
into all directions by some enemy or other event. From a more abstract point of
view, one has a nonequilibrium percolation problem.

This model is also closely related to the self--organized critical forest--fire
model (SOC FFM) \cite{dro92}, when occupied sites are equated with trees: For
$\rho < \rho_c$, the correlation length $\xi$ and the mean number of removed
sites per step, $S$, are finite, and all properties of the stationary state can
be found by looking at a section of the system of linear size $\ell$, with $\xi
< \ell \ll L$. In this section, there is no conservation of density, and the
dynamics can be characterized by a small 'tree growth' rate $p=S\rho
(\ell/L)^d/(1-\rho)$, and a 'lightning' rate $f=(\ell/L)^d$.  The tree growth
rate is the probability that a given empty site becomes occupied during one
step, and the lightning rate is the probability that a given site is 'struck
by lightning' (i.e., selected) per step, with the consequence that all
'trees' connected to this site 'burn down' (i.e., are removed). $S$
diverges in the limit $f/p \to 0$ as
\begin{equation}
S = p (1-\rho) / f \rho\, .
\label{barbara}
\end{equation}
In this limit, $\rho$ approaches the critical value $\rho_c$. The dynamics in
the small region of size $l$ are the same as for the SOC FFM, and the critical
exponents close to $\rho_c$ are therefore the same as those of the SOC FFM in
the limit $f/p \to 0$, when the $f/p$--dependence is translated properly into a
$\rho$--dependence (simulation results of the SOC FFM can be found in e.g.
\cite{hen93,gra93,chr93,cla94,hon96}).  Choosing $\rho$ as the control
parameter instead of $f/p$, allows us to study the FFM beyond the critical
point. This was in fact our original motivation to introduce the model studied
in this paper.

In the following three sections, we discuss in detail the subcritical, the
critical, and the supercritical behavior of the model.  Unless stated
otherwise, the considered system is a two--dimensional square lattice.

\section{Subcritical Phase and Approach to the Critical Point}
\label{subcritical}

First, we discuss the parameter region $\rho < \rho_c$, where the system has a
cutoff in cluster size that is independent of the system size $L$.  Let $S$ be
the mean number of sites removed from the system in one time step, without
taking into account the refilling of sites.  For very small densities, there
exist only very small clusters, and most clusters will consist of only one
site.  The process of removing clusters (in this case mostly single sites) and
refilling sites at random does not change this situation, i.e. $S$ is of the
order of one, and the occupied sites remain randomly distributed, as for a
percolation system at small densities.

With increasing density, $S$ increases also. Removing the chosen cluster and
refilling its sites at random into the system introduces fluctuations in the
local density of occupied sites, because the removed cluster leaves behind a
'hole' and the refilled sites increase the density in the rest of the system
to a value larger than $ \rho$. The critical density $\rho_c$, where $S$
diverges, and the critical exponents close to $\rho_c$, are therefore different
from their values in percolation theory.

We define the usual quantities that are investigated in percolation systems
(for an introduction to percolation, see \cite{sta92}): The number density of
clusters of occupied sites of size $s$ will be denoted by $n(s)$.  Near a
critical density $\rho_c$, $n(s)$ is expected to behave like a power law
\begin{equation}
\label{clusters}
n(s) \propto s^{-\tau} {\cal C}(s / s_{\rm{max}}),
\end{equation}
where ${\cal C}$ is a scaling function and $s_{\rm{max}} \propto |\rho_c -
\rho|^{-1/\sigma}$.  The average cluster size $S$ is defined by
\begin{equation}
S = \frac{\sum_{s=1}^\infty s^2 n(s)}{\sum_{s=1}^\infty s n(s)}
\end{equation}
and is expected to diverge like
\begin{equation}
S \propto |\rho_c - \rho|^{-\gamma}.
\label{smean}
\end{equation} 
The correlation length $\xi$ is defined as the root mean square distance
between occupied sites on the same cluster, averaged over all clusters, which
leads to
\begin{equation}
\xi^2 = \frac{\sum_{s=1}^\infty R^2(s) s^2 n(s)}{\sum_{s=1}^\infty s^2 n(s)}.
\end{equation}
Near a critical point, $\xi$ is expected to diverge like
\begin{equation}
\label{corrlength}
\xi \propto |\rho_c - \rho|^{-\nu}.
\end{equation}
The radius of gyration $R$ of a cluster grows with its size $s$ like
\begin{equation}
\label{radius}
R(s) \propto s^{1/d_f}
\end{equation}
with the fractal dimension $d_f$. These critical exponents are related via the
scaling relations $1/\sigma = \gamma / (3 - \tau) = d_f \nu$.  Finally, the
strength of an infinite cluster is denoted by $P$.  In percolation, $P$ follows
a power law
\[
P \propto (\rho - \rho_c)^\beta
\]
above the critical point.

In our simulations, we found $\rho_c\simeq 40.8\%$, $\tau = 2.15(3)$, $d_f =
1.96(2)$, $\nu = 1.20(5)$, and $\gamma = 2.09(5)$. The values of these
exponents as well as the value of the critical density $\rho_c$ are different
from percolation theory, and they are identical with the corresponding values
of the SOC FFM. The $(\rho_c - \rho)$--dependence of our model can be
translated into the $f/p$--dependence of the FFM using Eq.~(\ref{smean}) and
Eq.~(\ref{barbara}), giving
\[
\rho_c - \rho \propto (f/p)^{1/\gamma}.
\]
Thus, our values of $\sigma$ and $\nu$ can be calculated from those of the FFM
by multiplication with $\gamma$. $\tau$ and $d_f$ are exponents related to the
cluster size $s$, so no multiplication with $\gamma$ is necessary.  The
exponent $\beta$ vanishes, as we shall see in the next section.

\begin{figure}
\vskip 2cm
\centerline{\psfig{file=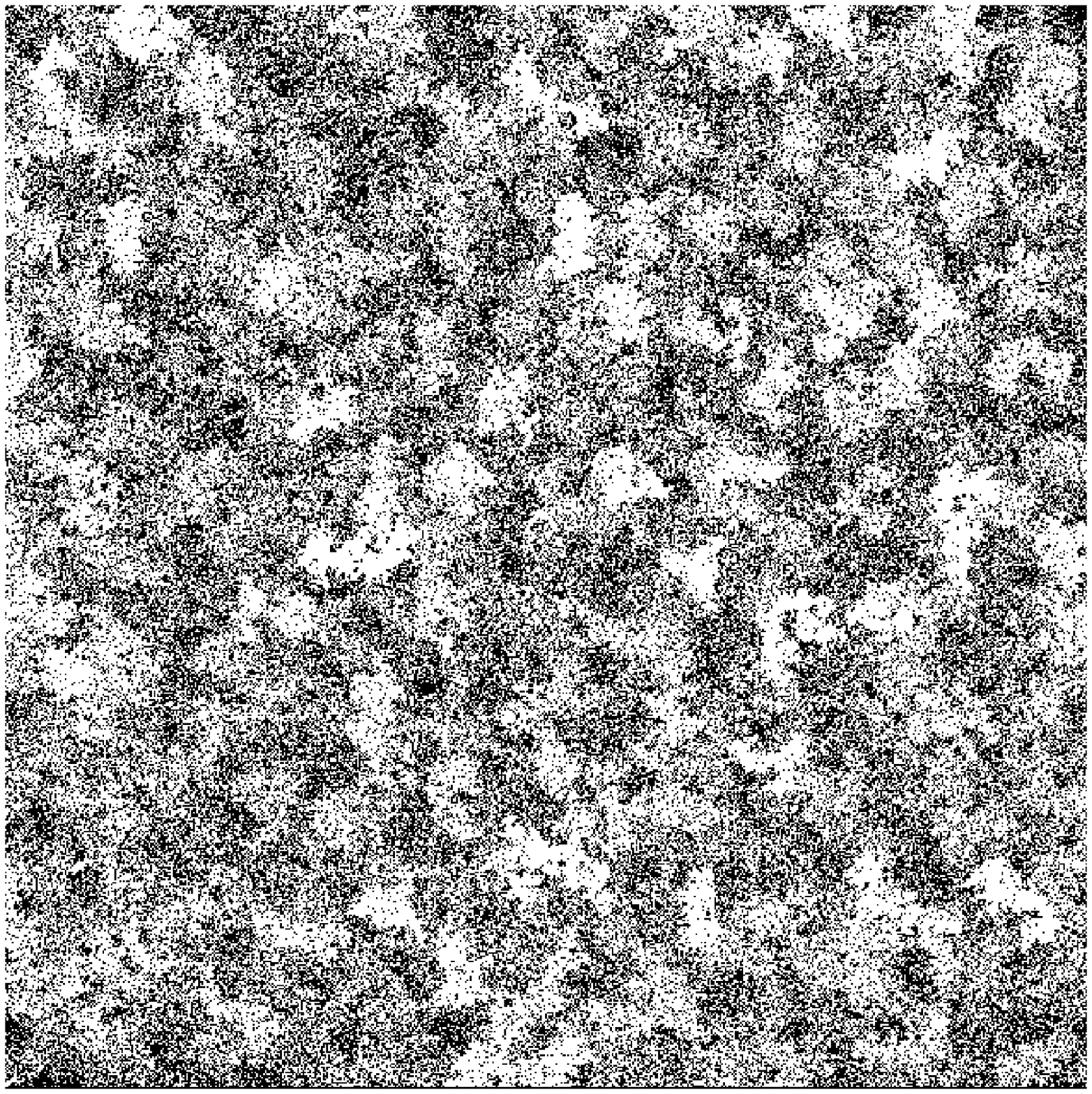,height=4in,angle=-90}} 
\vskip 1cm
\caption{Stationary state at $\rho = 39.3\%$ 
  and $L = 1024$ (periodic boundary conditions, square lattice with nearest
  neighbors, occupied sites are black, empty sites are white)}
\label{smallpatches39}
\end{figure}

The above--mentioned fluctuations in the local density of occupied sites can
easily be seen by looking at a snapshot of the system for densities close
enough to the critical density.  A typical stationary state for the density
$\rho = 39.3\%$ and system size $L = 1024$ is shown in
Fig.~\ref{smallpatches39}. One can see that the system consists of a large
number of regions with different and rather homogeneous density.  The typical
size of these 'patches' does not depend on the system size $L$, provided that
$L$ is large enough.  Many properties of the model can be understood by
describing the system in terms of these patches of homogeneous density of
occupied sites. For small average patch size (like in
Fig.~\ref{smallpatches39}), it is not always possible to assign a given site
unequivocally to a certain patch. This changes, however, when the critical 
density is approached, where the mean patch size is larger and the patch 
boundaries become sharper. 

For large system size $L^2 \gg S$, only a few sites are occupied in a given
patch per time step, and the density $\rho(t)$ in a patch evolves continuously
according to $\rho'(t) = p (1 - \rho(t))$ with some growth rate $p$, which
leads to
\begin{equation}
\label{macarena}
\rho(t) = 1 - (1 - \rho^{\rm{after}}) \exp(-pt) \, .
\end{equation}
Here, time is measured since the removal event that produced that patch,
leaving behind a small density of occupied sites $\rho(0) \equiv
\rho^{\rm{after}}$. We also define the mean density $\rho^{\rm{before}}$ of a
patch just before its spanning cluster is removed, and the time $T$ that it
takes to increase the density from $\rho^{\rm{after}}$ to $\rho^{\rm{before}}$.
We easily derive $T = (1/p) \ln((1 - \rho^{\rm{after}}) / (1 -
\rho^{\rm{before}}))$.  The average density of a patch is
\begin{eqnarray}
\rho & = & \frac{1}{T} \int_0^T dt (1 - (1 - \rho^{\rm{after}})
e^{-pt}) \nonumber \\ & = & 1 - \frac{\rho^{\rm{before}} -
\rho^{\rm{after}}}{\ln( \frac{1 - \rho^{\rm{after}}}{1 -
\rho^{\rm{before}}})}.  \label{ines}
\end{eqnarray}

\begin{figure}
\psfig{file=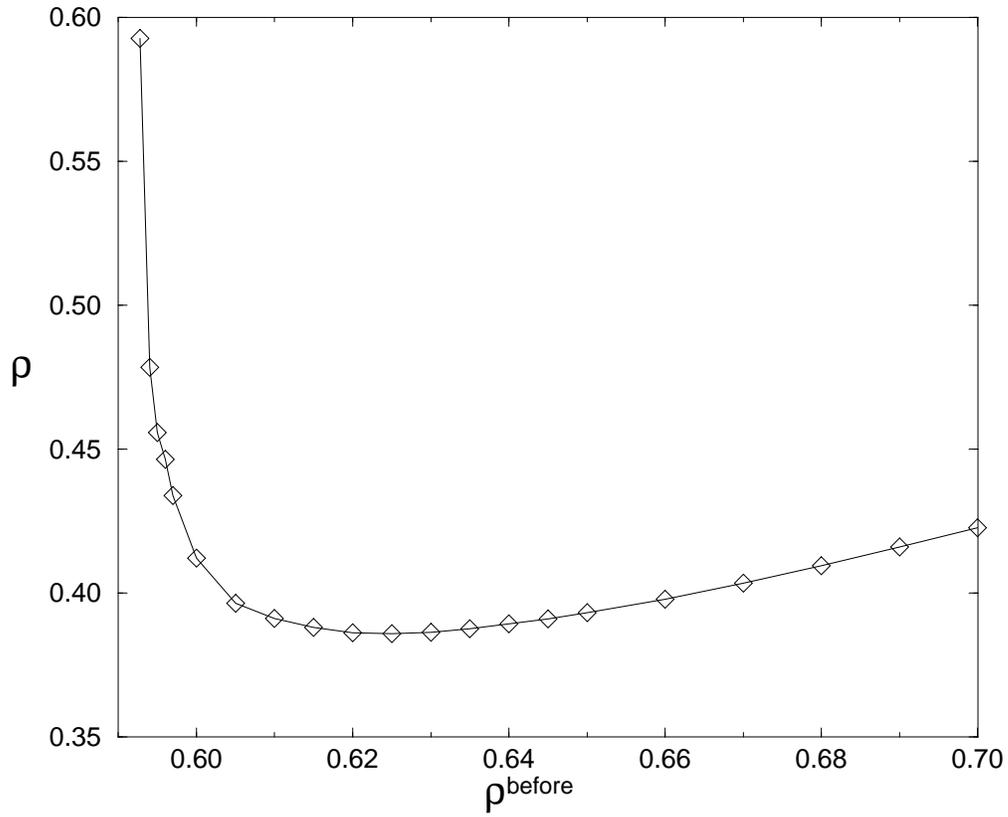,height=5in,angle=-90} 
\vskip 1cm
\caption{The average density $\rho$ for various values of
  $\rho^{\rm{before}}$ between $59.3\%$ and $70\%$. The minimum is at
  $\approx 62.5$.}
\label{strength}
\end{figure}

We measured the average values $\rho^{\rm{before}} \approx 62.5\%$ and
$\rho^{\rm{after}} \approx 7.8\%$, similar to the values found in
\cite{cha95a,hon96}. Interestingly, the same values will play an important role
in the critical phase and in the striped phase discussed in the following two
sections.  With the measured values of $\rho^{\rm{before}}$ and
$\rho^{\rm{after}}$, we obtain as average density of one region $\rho =
39.2\%$.  For large enough system size $L^2$ and neglecting the interactions
between the different regions, the time average of the overall density $\rho$
would also be $\rho = 39.2\%$.  In a real system with interacting regions, the
mean density is in general different due to the following two mechanisms: (i)
Temporal oscillations in patch size: During the growth process of the density
of a patch from $\rho^{\rm{after}}$ to $\rho^{\rm{before}}$, all the
neighboring patches have one removal process on an average. During each of
those removals, also some sites of the central patch are removed, because patch
boundaries are not cluster boundaries. This leads to a shrinking of the area of
the original patch while its density increases, reducing the mean density from
the above calculated value. (Of course, the original patch size is ultimately
restored when the spanning cluster of the central patch, and with it some
finite clusters of neighboring patches, are finally removed.)  The relative
shrinking of the area is stronger for smaller mean size of removed patches,
since the number of sites removed from the central patch depends on the length
of its boundary. For large mean size of removed patches, the ratio of initial
patch size to final patch size approaches 1, i.e. $\rho = 39.2\%$, which
explains why the density is smaller than the critical density when the mean
size of removed patches (of the order $S$) is finite.  (Fluctuations in the
size of removed patches are complicated to consider, and do not affect our main
conclusions.)  (ii) The critical density itself is larger than the above value
$\rho = 39.2\%$, because the density at which a removal event takes place
fluctuates around its mean value $\rho^{\rm{before}}$.  Assuming a homogeneous
distribution of occupied sites within a region, we measured the mean density
$\rho$ for various densities $\rho^{\rm{before}}$ between $59.3\%$ and $70\%$.
The result is shown in Fig.~\ref{strength}. One can see that this function has
a minimum for the measured average value of $\rho^{\rm{before}} \approx
62.5\%$.  Any fluctuation will therefore lead to an increase in the mean value
of $\rho$. The fact that the density is so close to its minimum indicates a
tendency of the system to maximize energy dissipation.  To obtain the order of
magnitude of the shift in $\rho_c$ due to fluctuations, we calculated
numerically the mean density in the simulated interval $[p_c; 0.70]$ of
$\rho^{\rm{before}}$. The result is $40.5(3)\%$, which is close to the correct
critical density $\rho_c\simeq 40.8\%$, indicating that the fluctuations in
$\rho^{\rm{before}}$ can indeed induce the observed shift in the density.

These considerations show that the critical state of our model can to good
approximation be interpreted as a combination of percolation systems of
different densities, as also suggested in \cite{hon96}.  The cluster size
distribution $n(s)$ is therefore the superposition of the cluster size
distributions of all the patches with their different densities between
$\rho^{\rm{after}}$ and $\rho^{\rm{before}}$, distributed according to
Eq.~(\ref{macarena}).

Let us first consider the one--dimensional system. There, the percolation
threshold is one, and even in dense regions all clusters are finite percolation
clusters. Removal of a cluster leaves behind a string of empty sites. The
system is therefore composed of strings of size $S$ of different densities that
represent different stages in the growth process of an empty string until it is
completely filled. A string of density $\rho$ contains a cluster of size $s$
with probability $\rho^s (1 - \rho)^2$.  If the system size $L$ is large
enough, the time average equals the ensemble average, and the cluster size
distribution should be given by
\begin{eqnarray*}
  n(s) & \propto & \int_0^\infty dt \ \rho(t)^s (1 - \rho(t))^2
  \nonumber \\ & = & \int_0^\infty dt \ (1-e^{-p t})^s (e^{-p t})^2
  \nonumber \\ & \propto & \int_0^\infty dt \ \sum_{k = 0}^s {s
    \atopwithdelims() k} (-1)^k e^{-kt} e^{-2t} \nonumber \\ & = &
  \sum_{k = 0}^s {s \atopwithdelims() k} (-1)^k {1 \over 2 + k}
  \nonumber \\ & = & \frac{1}{(s+1) (s+2)} \nonumber \\ & \approx &
  s^{-2} \, \, \text{for large } s\, ,
\end{eqnarray*}
in agreement with the exact calculation of \cite{dro93}. Similarly, the size
distribution of hole clusters is found to be
\begin{eqnarray*}
  h(s) & \propto & \int_0^\infty dt \ \rho(t)^2 (1-\rho(t))^s
  \nonumber \\ & \propto & \frac{1}{s (s+1) (s+2)} \nonumber \\ &
  \approx & s^{-3} \, \, \text{for large } s\, ,
\end{eqnarray*}
again in agreement with exact results \cite{pac93,dro93}.

\begin{figure}
\psfig{file=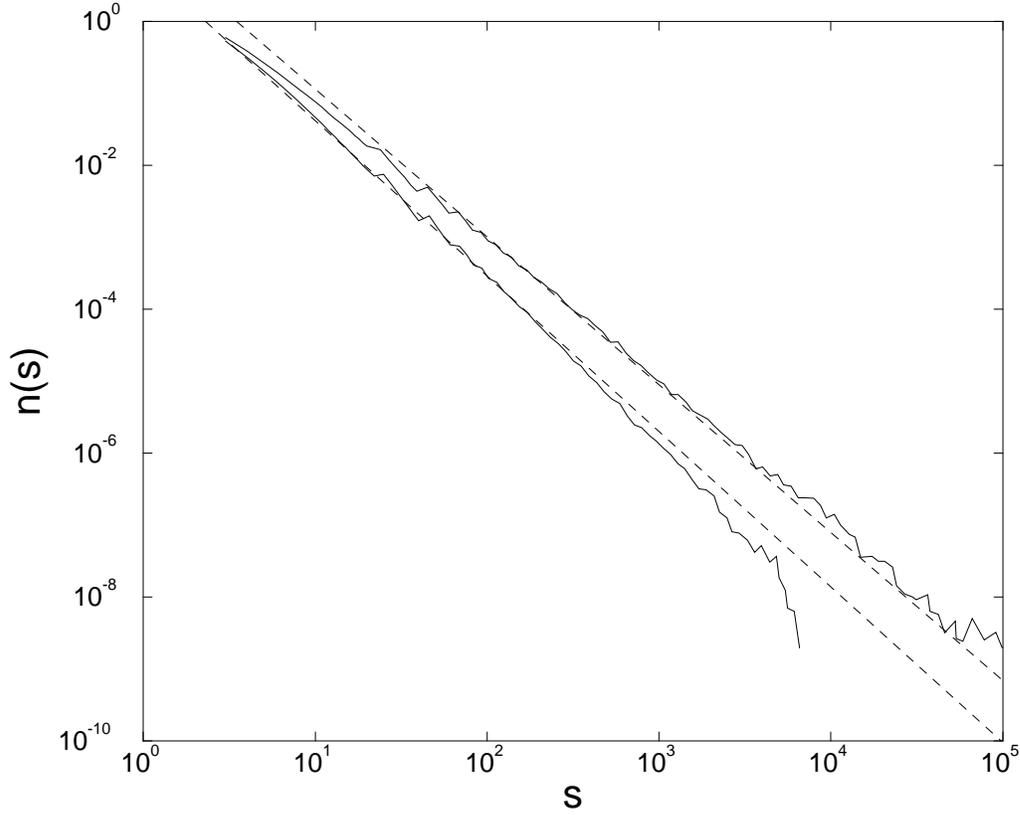,height=5in,angle=-90} 
\vskip 1cm
\caption{Upper curve: cluster size distribution $n(s)$ of a homogeneous system
  at the percolation threshold $p_c \approx 59.3\%$. Lower curve: cluster size
  distribution $n(s)$ of a system with densities between $\rho^{\rm{after}}$
  and $\rho^{\rm{before}}$, distributed according to Eq.~(\protect
  \ref{macarena}). Both systems have size $L = 2048$. The dashed straight lines
  have slope 2.06 and 2.15, respectively.}
\label{revelation}
\end{figure}

For dimensions higher than one, unfortunately, these calculations can not be
carried out. There, the number of neighbors $t$ of a cluster of size $s$ does
not only depend on $s$, but also on the shape of the cluster, and
the number of different clusters with a given size $s$ and a given perimeter
$t$ is not known exactly.
Furthermore, the integration over $t$ does not go from zero to infinity, since
the lowest and highest densities are not 0 and 1, respectively. 
We can, however, determine numerically the cluster size distribution of a 
combination of percolation systems of different densities. We measured
$n(s)$ in 20 homogeneous systems with
$L = 2048$ and different densities between $\rho^{\rm{after}}$ and
$\rho^{\rm{before}}$, distributed according to Eq.~(\ref{macarena}). 
The total $n(s)$ of all systems $\propto \sum_{i=1}^{20} n_i(s)$ is plotted in
Fig.~\ref{revelation}, together with the cluster size distribution of a
percolation system at $\rho = p_c \approx 0.593$.  The exponent $\tau \approx
2.15$ for a combination of percolation systems is indistinguishable from the
cluster size distribution exponent of our model, showing that the latter is
indeed generated by a superposition of percolation systems.
Fig.~\ref{revelation} does not show the bump observed in our simulations for
large $s$.  For large cluster sizes, the size distribution of patches with
density above the percolation threshold affects the cluster size distribution.
This effect is not present in Fig.~\ref{revelation}, where there are no
different patch sizes. It remains an open question whether the exponent $\tau$
remains the same for much larger values of $s$ than those accessible to
simulations. At very large scales, the emerging dynamics of patches might
become an important factor in determining the cluster size distribution.

Another conclusion that can be drawn, is that $\tau$ has to be always larger
than or equal to the $\tau$ of the corresponding percolation problem.  $n(s)$
results from a superposition of many percolation systems, so that apart from
the critical percolation system, there exist also many systems with finite
cutoff in the cluster size distribution. These systems increase the weight of
the small clusters, and the slope of $n(s)$ decreases, resulting in a larger
$\tau$. For all dimensions and lattice types investigated so far (see for
example \cite{cla94}), this was true.

\section{The Critical Phase, Hyperscaling, and Finite-Size Scaling}
\label{criticalregion}

\begin{figure}
\vskip 1cm
\centerline{\psfig{file=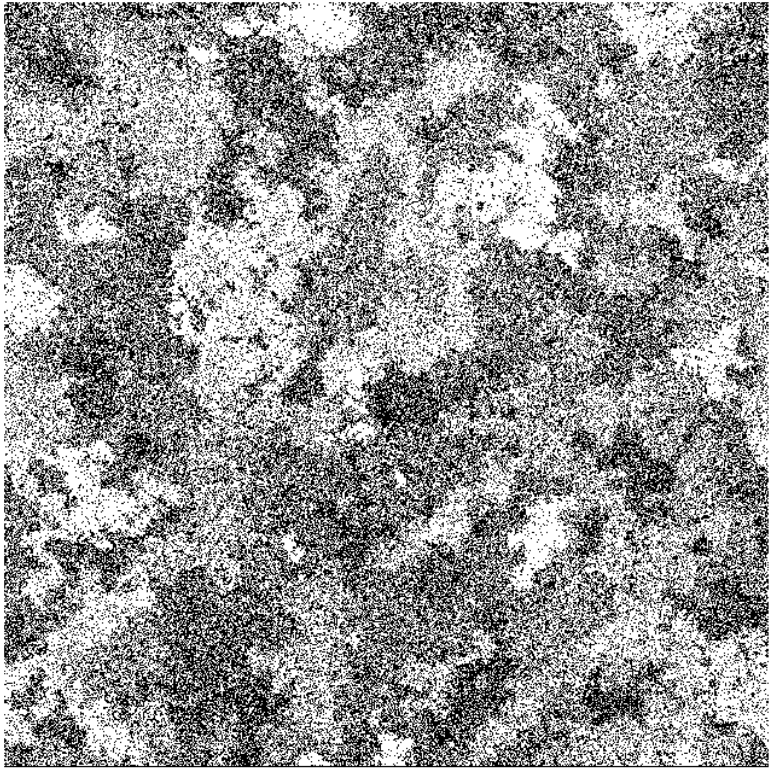,height=4in,angle=-90}} 
\vskip 1cm
\caption{Stationary state at the critical density $\rho_c \approx 40.8\%$ 
  and $L = 4096$ (periodic boundary conditions, square lattice with nearest
  neighbors, occupied sites are black, empty sites are white)}
\label{criticalpoint}
\end{figure}

\begin{figure}
\centerline{\psfig{file=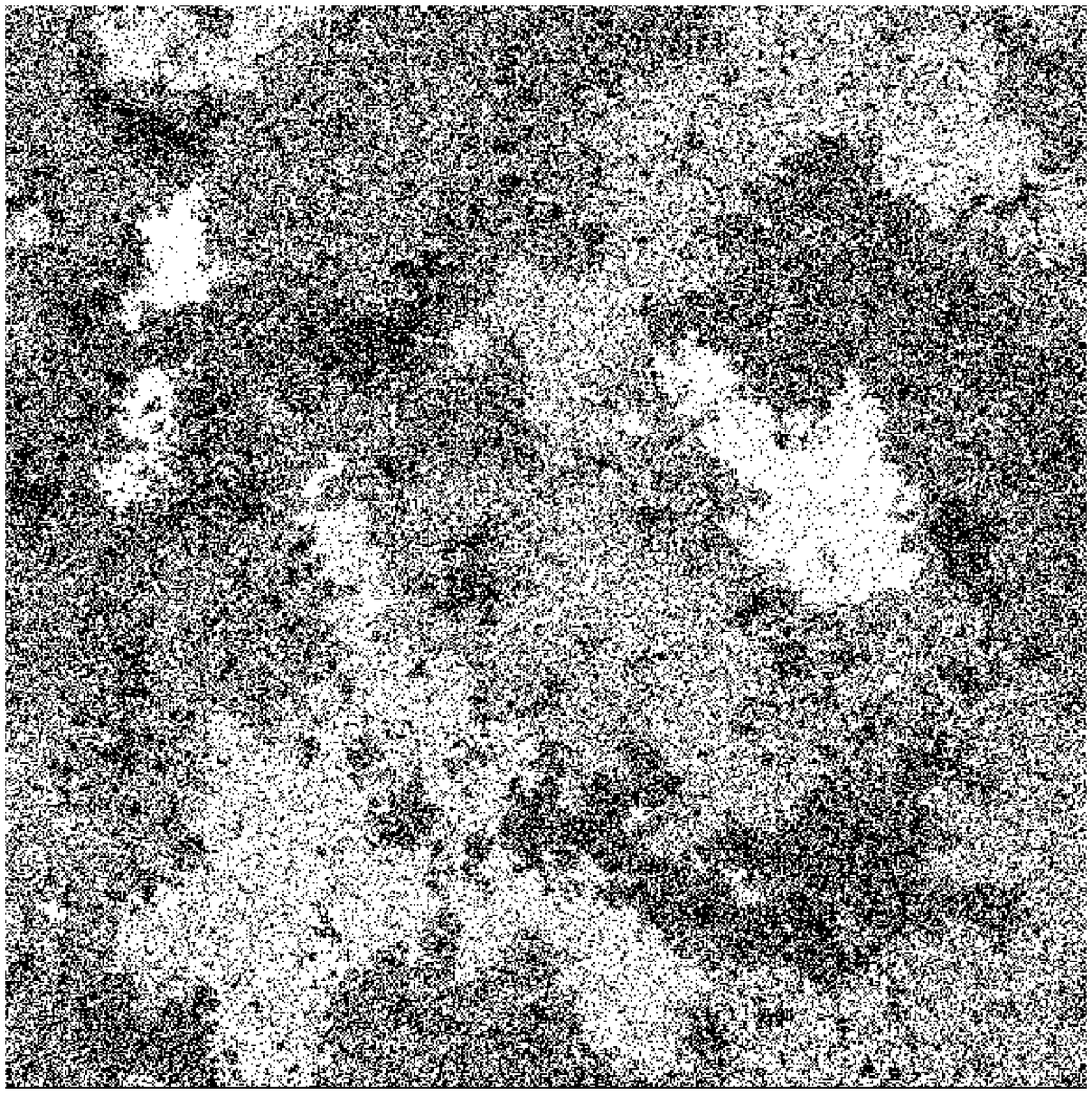,height=4in,angle=-90}} 
\vskip 1cm
\caption{Stationary state at $\rho = 43\%$ 
  and $L = 4096$ (periodic boundary conditions, square lattice with nearest
  neighbors, occupied sites are black, empty sites are white)}
\label{largepatches43}
\end{figure}

For $\rho > \rho_c$, one might expect the appearance of an infinite cluster
which spans the whole system, as in percolation theory.  However, the values of
the critical exponents for $\rho \lesssim \rho_c$ already indicate that the
situation in our system is very different from percolation. In contrast to
percolation, the hyperscaling relation $d = d_f (\tau - 1)$ is violated, which
means that not every part of the system contains a spanning cluster at
criticality \cite{hen93}. This can also be seen easily by looking at the
snapshot of a system at $\rho \approx \rho_c$ in Fig.~\ref{criticalpoint}.  In
addition to large patches with high density, there exist also large patches
with lower densities. Most of the patches have densities below the percolation
threshold and contain many finite clusters. The patches with high densities
(between $p_c$ and $\approx 62.5\%$) contain not only finite, but also
'infinite' (i.e., spanning) clusters.  In contrast to ordinary percolation,
there is no homogeneously distributed set of large clusters that could join at
$\rho = \rho_c$ to form the infinite cluster that spans the whole system.
Rather, besides the largest cluster, the system contains many other large
clusters that represent different growth stages of the largest cluster itself.
The critical behavior of our model occurs over a finite interval $[\rho_c^{(1)}
(= \rho_c \approx 40.8\%), \rho_c^{(2)} (\approx 43.5\%)]$, where the cutoff in
cluster size $s_{\rm{max}}$ diverges, but slower than $L^2$. The correlation
length diverges slower than $L$. Since it is not possible to define a truly
infinite cluster like in percolation in this phase, all clusters in the system
contribute to the defining equations of $s_{\rm{max}}$, $\xi$ and $S$. We
find
\[
  s_{\rm{max}} \propto L^{\phi_1}, \xi \propto L^{\phi_2}, S \propto
  L^{\phi_3}
\]
with $\rho$--dependent exponents $\phi_{1,2,3}$, while $\tau$ and $d_f$ remain
unchanged. Tab.~\ref{fsexpo} shows the values of the exponents for different
densities.  Fig.~\ref{largepatches43} shows a snapshot of the system for $\rho
= 0.43$.

\begin{figure}
\psfig{file=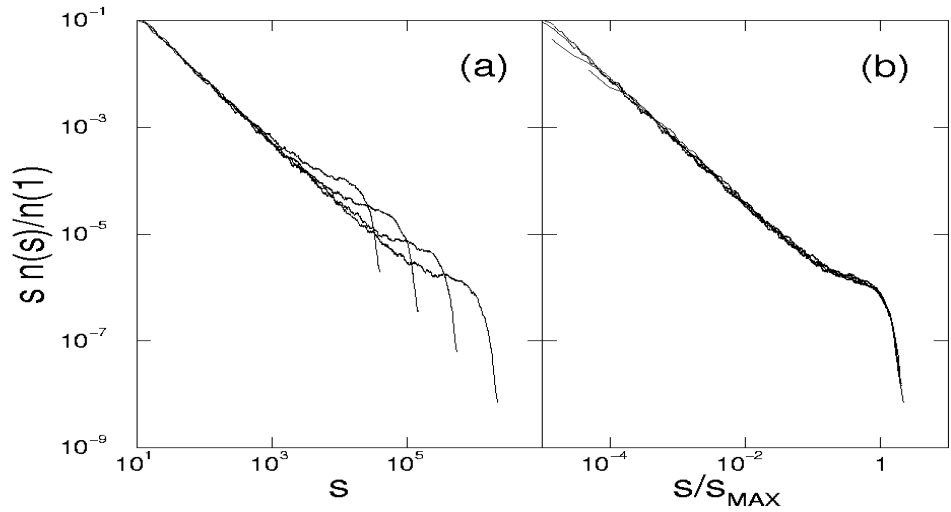,height=3in,width=5in,angle=-90}
\vskip 1cm
\caption{Normalized size distribution of clusters for 
  $\rho = 0.43$ and $L$ = 512, 1024, 2048, 4096, (a) before and (b) after
  rescaling.}
\label{scaling43}
\end{figure}

Fig.~\ref{scaling43} shows the size distribution of clusters $s n(s)$ for
different system sizes at fixed $\rho$ (a) before and (b) after rescaling.
With Eqs.~(\ref{clusters}) - (\ref{radius}) one can derive the scaling
relations $\phi_1 = d_f\phi_2$ and $\phi_3 = (3 - \tau) \phi_1$, which are well
confirmed by the simulations.  For $\rho \to \rho_c^{(2)}$, $\phi_{1,2,3}$
approach the values $d_f$, 1, and $1/\nu$, respectively.

The latter values are those that one would already have expected at $\rho =
\rho_c^{(1)}$ from finite-size scaling theory. In a critical system, a quantity
$X$ that scales as $|T - T_c|^{-\chi} \propto \xi^{\chi / \nu}$ in an infinite
system, in a finite system is expected to obey the scaling form
\[
  X(L, \xi) = \xi^{\chi / \nu} \cdot \hat X(\xi / L)
\]
with
\[
  \hat X(x) = \left\{
\begin{array}{ll} 
  \text{const} & \text{for } x \ll 1 \\ x^{-\chi / \nu} & \text{for }
  x \gg 1
\end{array} 
\right.
\]
such that for $\xi \ll L$, $X \propto |T - T_c|^{-\chi}$ like in the infinite
system, and for $L \ll \xi$, $X \propto L^{\chi / \nu}$.  The standard
finite-size scaling exponents of $s_{\rm{max}}$, $\xi$ and $S$ would
therefore be $1 / \nu \sigma = d_f$, $\nu / \nu = 1$, and $1 / \nu$.

The main difference between our system and percolation concerning finite-size
scaling is that in percolation a system of length $L_{\rm{large}}$ without
finite-size effects can be generated by putting together systems with
finite-size effects of length $L_{\rm{small}}$. In our model, a
reorganization takes place when smaller systems are put together, since the
smaller systems now can have fluctuations in their number of occupied sites
that they could not have when they were isolated. Due to the inhomogeneous
nature of the stationary state, these fluctuations are very large.

\begin{figure}
\psfig{file=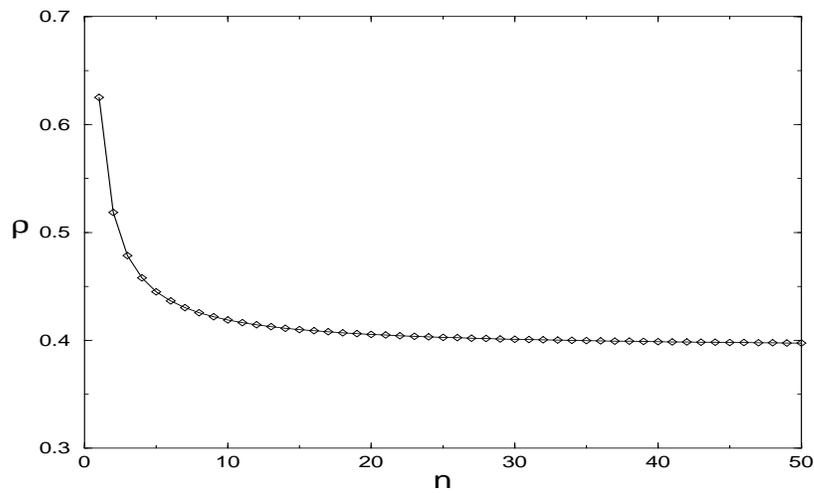,height=3in,width=5in,angle=-90}
\vskip 1cm
\caption{$\rho$ as function of the number of patches $n$ with densities between
  $\rho^{\rm{after}}$ and $\rho^{\rm{before}}$.}
\label{coarsened}
\end{figure}

We obtain a deeper understanding of the critical behavior of our model when we
describe the system again in terms of patches of different densities.  In the
previous section, we found that the density within a patch grows continuously
for $\rho<\rho_c$ and for sufficiently large system size.  For not sufficiently
large system size, and for $\rho > \rho_c$, however, the finite system size
leads to discontinuous jumps in the density within a patch. The non-continuous
version of Eq.~(\ref{ines}) is (see Eq.~(\ref{noncon}) below)
\[
  \rho = 1 - \frac{\rho^{\rm{before}} -
    \rho^{\rm{after}}}{n(((1-\rho^{\rm{before}})/
(1-\rho^{\rm{after}}))^{-1/n}-1)}.
\]
for the density of a system that takes $n$ time steps to go from
$\rho^{\rm{after}}$ to $\rho^{\rm{before}}$.  In Fig.~\ref{coarsened},
$\rho$ is plotted as function of the number of patches $n$. One can see that a
decrease in $n$ raises the mean density. (Of course, this conclusion can also
be obtained from an analytical calculation.) This effect supplements the two
other mechanisms that affect the mean density presented in the previous
section. For fixed $\rho$ and changing $L$, these three mechanisms have to be
in balance with each other. While the step size decreases with increasing $L$,
tending to decrease the density, the size of the large patches and the
fluctuations in $\rho^{\rm{before}}$ increase, cancelling the effect of the
smaller step size on the density.

So far, we have not yet discussed the possibility of fusion and splitting of
patches. As soon as two neighboring patches have a density above the
percolation threshold, they fuse and will be removed together. This effect must
be balanced by a mechanism that splits patches. As long as the density of a
patch is below the percolation threshold, neighboring removals move the
boundary of this patch inwards, and a splitting into two patches occurs when
opposite boundaries meet. A splitting can also occur when a finite cluster
connecting two opposite edges of the patch is removed before the density in the
patch reaches the percolation threshold. Since even large patches must split at
the same rate at which they fuse with neighboring patches, the shape of patches
cannot be round, but must be 'fingered', with necks of a width that does not
depend on the patch size.  The snapshots Figs.~\ref{criticalpoint} and
\ref{largepatches43} show the fingered structure.  It implies some
characteristic length scale, the 'finger thickness', besides the lattice
constant and the correlation length.  This scale is a function of the density
and the system size.  The existence of several length scales was also observed
in \cite{hon96}.

\section{Supercritical Phase}
\label{supercritical}

\subsection{Synchronized States}
\label{synchro}

At $\rho = \rho_c^{(2)}$, the correlation length finally becomes proportional
to the system size, and the system has an 'infinite' cluster that contains a
finite percentage (independent of $L$) of all occupied sites in the system.

\begin{figure}
\centerline{\psfig{file=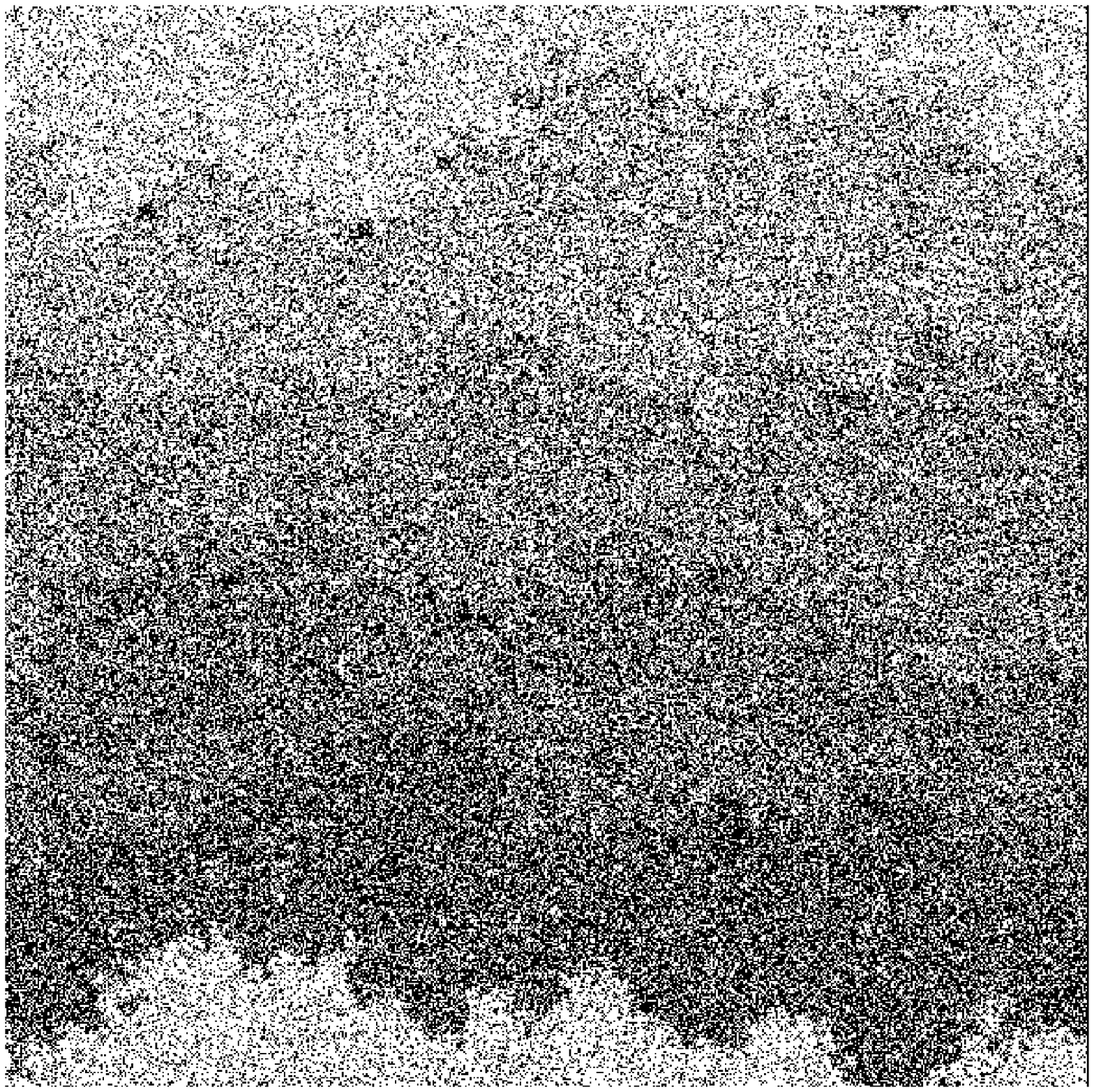,height=4in,angle=0}} 
\vskip 1cm
\caption{A stationary state with five stripes at $\rho = 0.45$ 
  and $L = 4096$ (periodic boundary conditions, square lattice with nearest
  neighbors, occupied sites are black, empty sites are white)}
\label{fivephases}
\end{figure}

Fig.~\ref{fivephases} shows a snapshot of the system for $\rho = 0.45$.  We see
a system with five homogeneous and equally large stripes with different
densities. The stripe with highest density is above the percolation threshold
and contains an infinite cluster (with non-zero strength $P$) as well as some
small clusters.  The other stripes are below the percolation threshold and,
consequently, contain only small clusters.  When one of the small clusters in
this state is removed, only few sites are redistributed, and the state of the
system remains essentially unchanged. When the infinite cluster is removed, a
large portion of all occupied sites in the stripe with the highest density are
redistributed all over the system.  The stripe which used to have the highest
density now has the lowest density, while the density of the other stripes has
increased. The values of the five densities are the same as before, except that
they are now associated with different stripes. If we measure time in units of
large redistributions of sites, the state of the system is periodic with period
5.

\begin{figure}
\centerline{\psfig{file=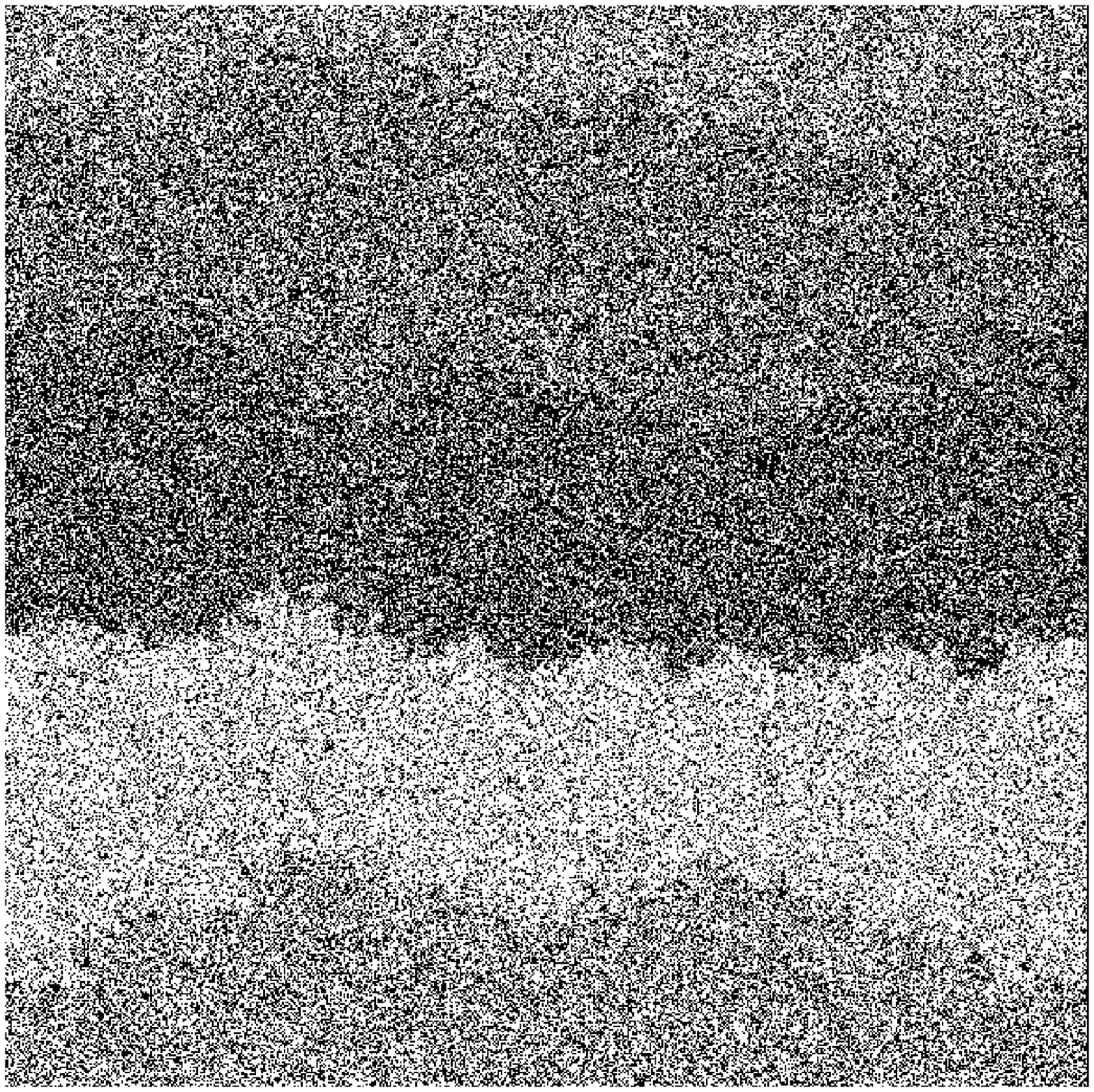,height=4in,angle=0}} 
\vskip 1cm
\caption{Stationary state with four stripes at $\rho = 0.46$ 
  and $L = 4096$ (periodic boundary conditions, square lattice with nearest
  neighbors, occupied sites are black, empty sites are white)}
\label{fourphases}
\end{figure}

\begin{figure}
\centerline{\psfig{file=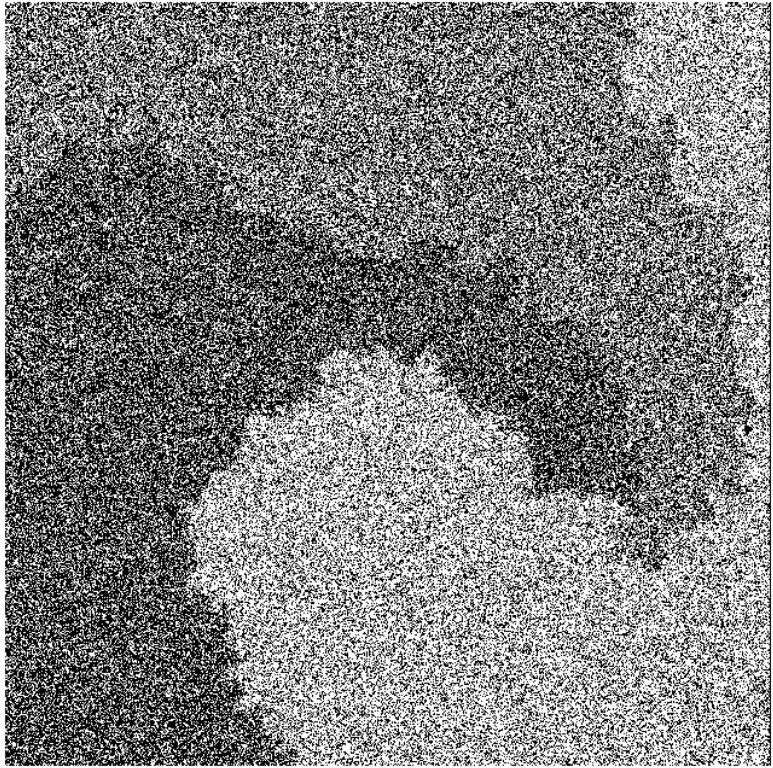,height=4in,angle=0}} 
\vskip 1cm
\caption{Stationary state with three regions of different density
  at $\rho = 0.50$ and $L = 2048$ (absorbing boundary conditions, square
  lattice with nearest neighbors, occupied sites are black, empty sites are
  white)}
\label{threephases}
\end{figure}

\begin{figure}
\centerline{\psfig{file=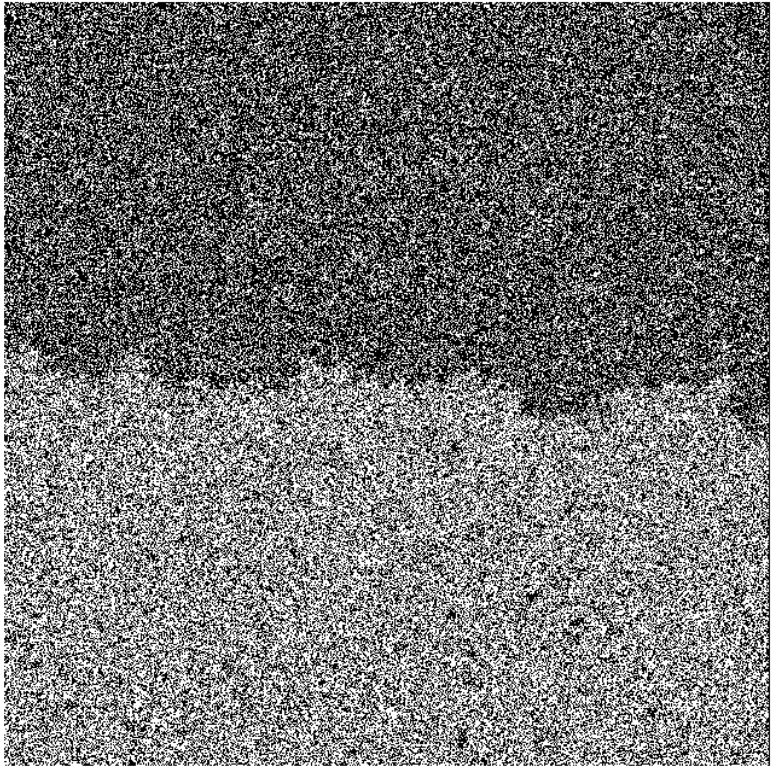,height=4in,angle=0}} 
\vskip 1cm
\caption{Stationary state with two stripes at $\rho = 0.55$ 
  and $L = 1024$ (absorbing boundary conditions, square lattice with nearest
  neighbors, occupied sites are black, empty sites are white)}
\label{twophases}
\end{figure}

\begin{figure}
\centerline{\psfig{file=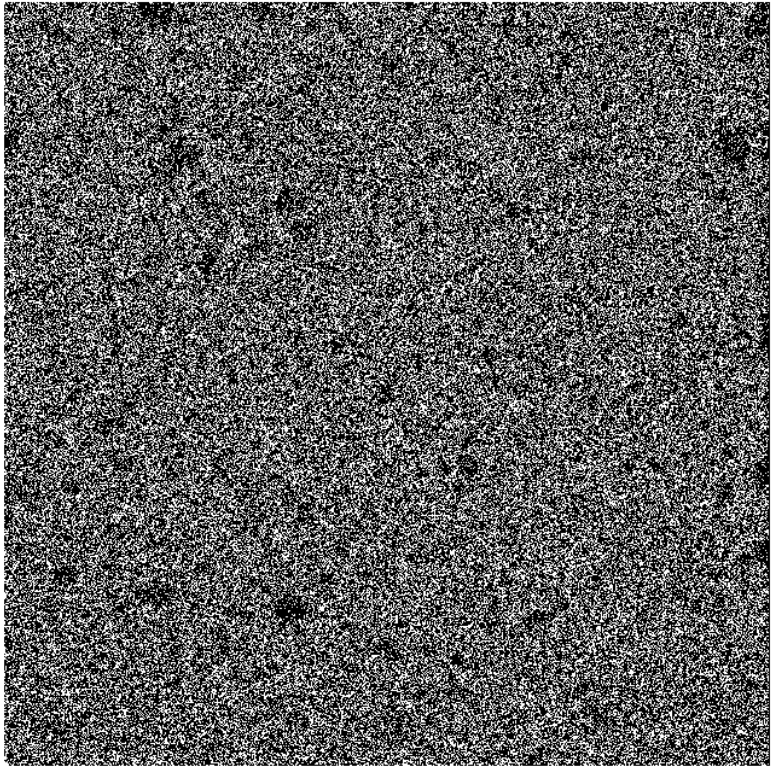,height=4in,angle=0}} 
\vskip 1cm
\caption{Stationary state at $\rho = 0.63$ 
  and $L = 1024$ (absorbing boundary conditions, square lattice with nearest
  neighbors, occupied sites are black, empty sites are white)}
\label{onephase}
\end{figure}

Increasing $\rho$, we find four (Fig.~\ref{fourphases}), three
(Fig.~\ref{threephases}), two (Fig.~\ref{twophases}), and finally one
'stripe' (Fig.~\ref{onephase}), where the infinite cluster spans the whole
system.  The spatial shape of stripes depends on lattice symmetry and boundary
conditions. In the case of a two-dimensional square or triangular lattice with
periodic boundary conditions, the system self-organizes into stripes with the
boundaries along one of the principal axes.  For absorbing boundary conditions,
the stripes are replaced by regions of a different shape (see
Fig.~\ref{threephases}).

To understand the occurrence of stripes with different densities, we consider
first a system with a very high density $\rho \lesssim 1$.  Since we are far
above the percolation threshold $p_c \approx 0.593$ for random site
percolation, the strength of the infinite cluster $P$ is close to $\rho$.  We
start with a random initial state.  In the first iteration step, we remove
either the infinite cluster, which consists of nearly all occupied sites in the
system, or one of the finite clusters, which, consequently, are very small.  In
the first case, only a few small clusters remain, and most of the occupied
sites are redistributed randomly in the system. In the second case, only a
small number of occupied sites are redistributed.  In both cases, however, the
state of the system changes little, and the new state is close to a completely
random state.

\begin{figure}
\psfig{file=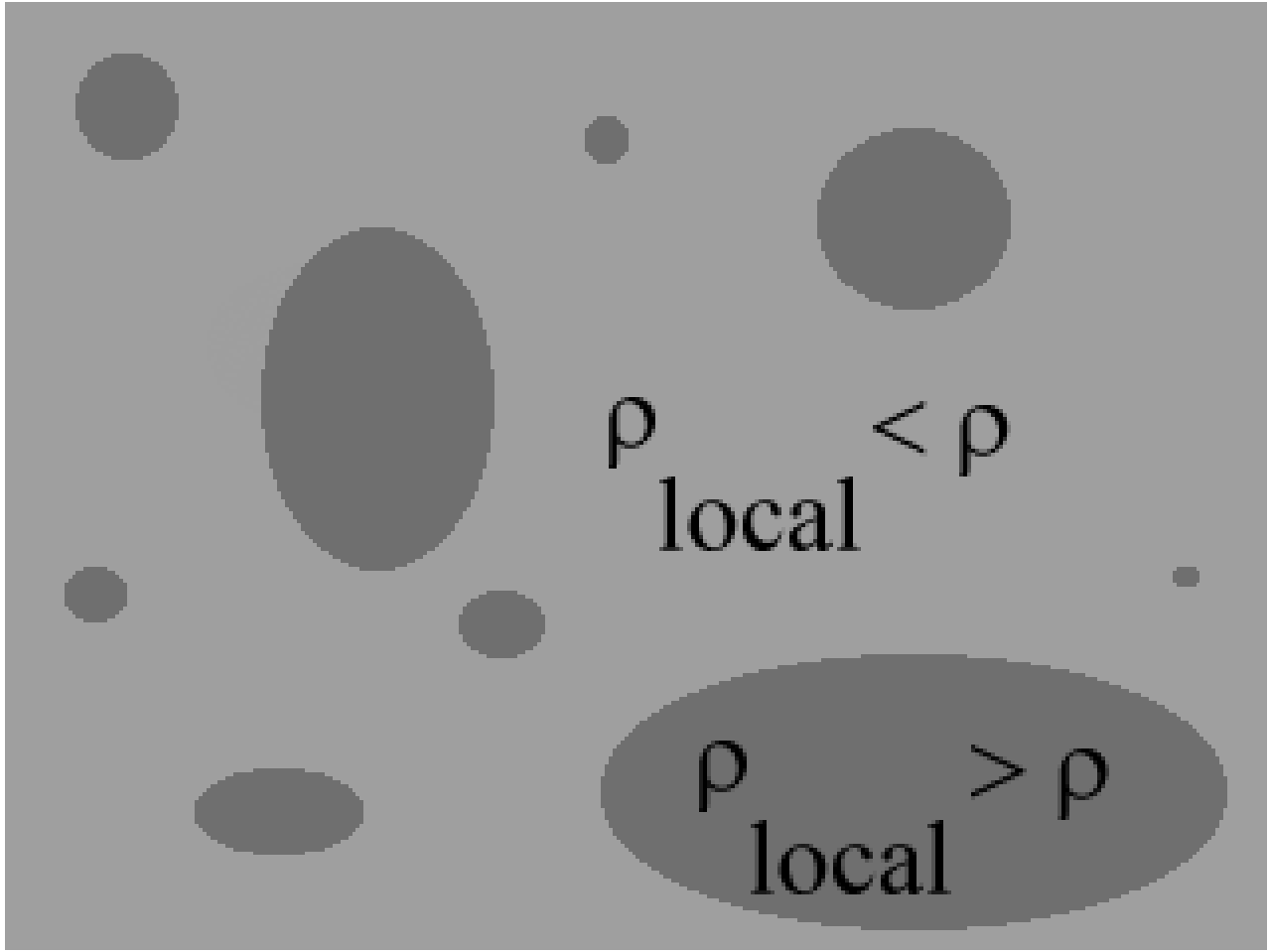,height=5in,angle=0} 
\vskip 1cm
\caption{Schematic picture of the stationary state at a density 
  just above $\rho^*$. Light (dark) grey regions have lower (higher) than
  average density.}
\label{rhoexplanation}
\end{figure}

If we now decrease $\rho$, the remaining clusters (after the removal of the
infinite cluster) become larger, and the density fluctuations increase.  When
the removed sites are refilled into the system, part of them are positioned in
or near these surviving clusters, where the density then will be larger than
the mean density. In the space between these clusters the density consequently
is lower than the mean density (see Fig.~\ref{rhoexplanation}).  If $\rho$
falls below a certain threshold $\rho^*$, this density between the surviving
clusters becomes smaller than the percolation threshold $p_c$. Then, there
exists no infinite cluster in the system after the first time step (or after a
few iterations), and the state with one stripe becomes unstable.  Evidently,
$\rho^* > p_c$.

The approximate value of $\rho^*$ can be derived by the following argument: The
number of empty sites after the removal of the infinite cluster is $L^2 (1 -
(\rho - P))$. Let this subset of the total lattice be denoted by $\cal D$. It
contains the area of the infinite cluster and the sites that have already been
empty before. In the next step, the $L^2 P$ occupied sites of the infinite
cluster are redistributed randomly among these empty sites, leading to a
density $\rho' = P / (1 - (\rho - P)) < \rho$ in $\cal D$.  In the vicinity of
the clusters that were left over the local density is now higher than before.
One can think of this state approximately as consisting of many compact
clusters embedded into a lower-density background (see also
Fig.~\ref{rhoexplanation}).  It is then obvious that the homogeneous state with
an infinite cluster stretching over the whole system cannot survive if $\rho' <
p_c$, since, in this case, the infinite network which links the finite
high--density regions is broken. On the other hand, if $\rho' > p_c$, the
situation is not fundamentally different from the case $\rho \lesssim 1$.
Fluctuations in the local density do exist, but they cannot get stronger with
time, since the high-density regions are themselves removed from the system
with high probability during the next few iteration steps.  Thus, we arrive at
the following implicit approximate equation for a threshold density $\rho^*$.
\begin{equation}
\label{rhostar}
\frac{P(\rho = \rho^*)}{1 - \rho^*} = \frac{p_c}{1 - p_c}.
\end{equation}
In the simulations, we found $\rho^* \approx 0.625$ for the square and 0.533
for the triangular lattice.  A measurement of the left hand site of
Eq.~(\ref{rhostar}) at $\rho = \rho^*$ and $L = 1024$ yielded $\approx 1.46$
for the square and $\approx 1.01$ for the triangular lattice.  The value of the
known right hand site is $\approx 1.46$ and 1, respectively. Although
Eq.~(\ref{rhostar}) is approximate, the agreement with simulation results is
very good.

If $\rho$ falls below $\rho^*$, the homogeneous phase becomes unstable, and the
system rearranges itself to a new stationary state that consists of two
homogeneous stripes with equal area and different densities $\rho_1$ and
$\rho_2$ ($\rho_1 > \rho_2$).  For $\rho \lesssim \rho^*$, we have $\rho_1 >
\rho^*$ and $\rho_2 < \rho^*$.  When the infinite cluster in this new state is
removed, the density of stripe 1 decreases to $\rho_3 \equiv \rho_1 - P$. Now
these sites are re-injected randomly into the system, such that stripe 1 is
filled up to $\rho_2$ and stripe 2 to $\rho_1$.  The net effect is that the two
stripes have changed their roles.

If we lower $\rho$ further, we will eventually reach the density $\rho$ for
which $\rho_1 = \rho^*$. At this point the two-stripe state becomes itself
unstable (because the highest density stripe becomes unstable) and reorganizes
into a three-stripe state, thereby increasing $\rho_1$ such that $\rho_1 >
\rho^*$ again. The dynamics of the three-stripe state is analogous to the
two-stripe state: When the infinite cluster is redistributed, the stripes
simply exchange their densities ($3 \to 2$, $2 \to 1$, $1 \to 3$).  With
decreasing density, the number of stripes $n$ increases further.

\begin{figure}
\psfig{file=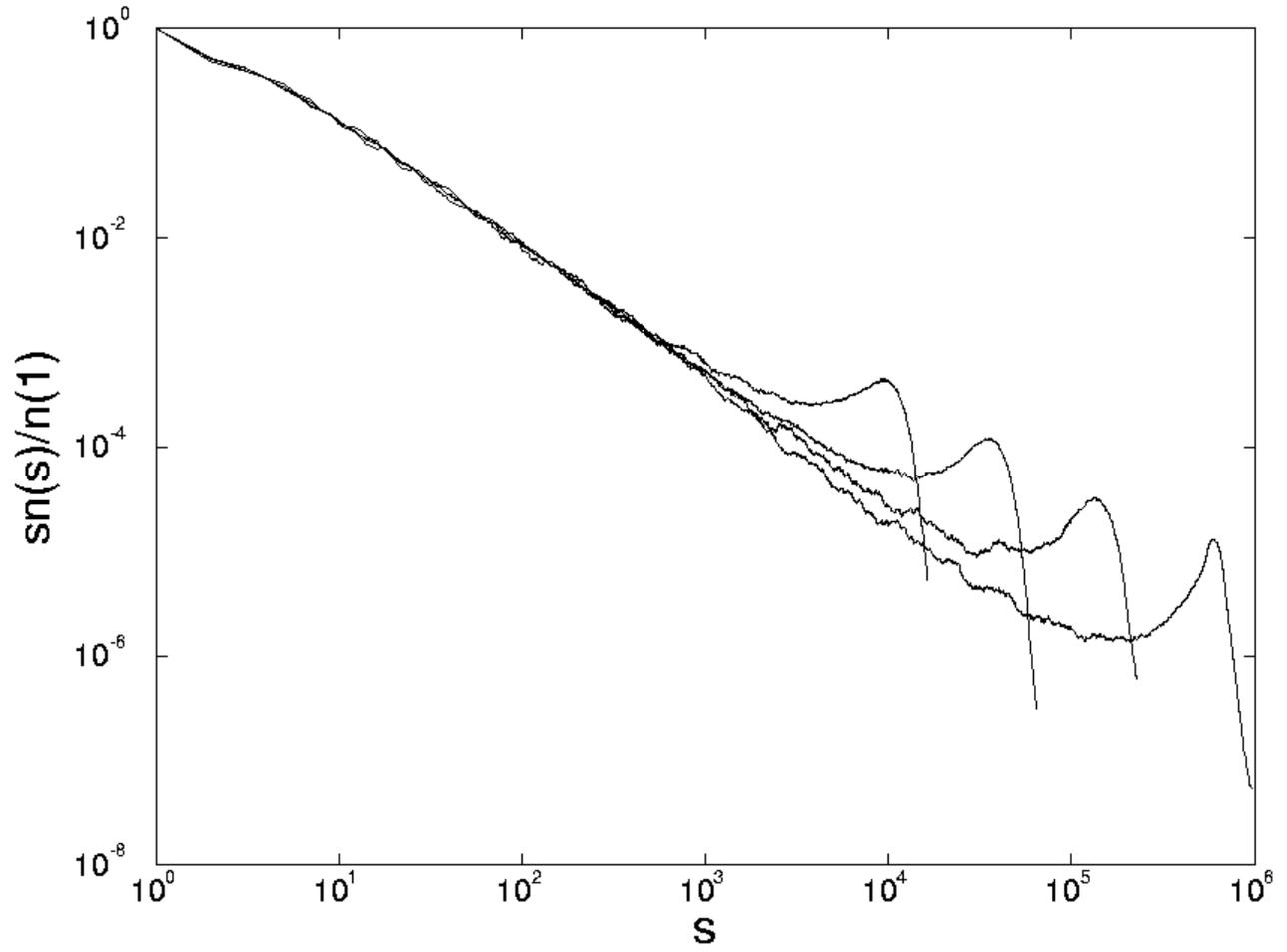,height=5in,angle=-90}
\vskip 1cm
\caption{Cluster size distribution of a system at $\rho = 0.46$ for various
  system sizes ($L = 512, 1024, 2048, 4096$).}
\label{finsize}
\end{figure}

\begin{figure}
\psfig{file=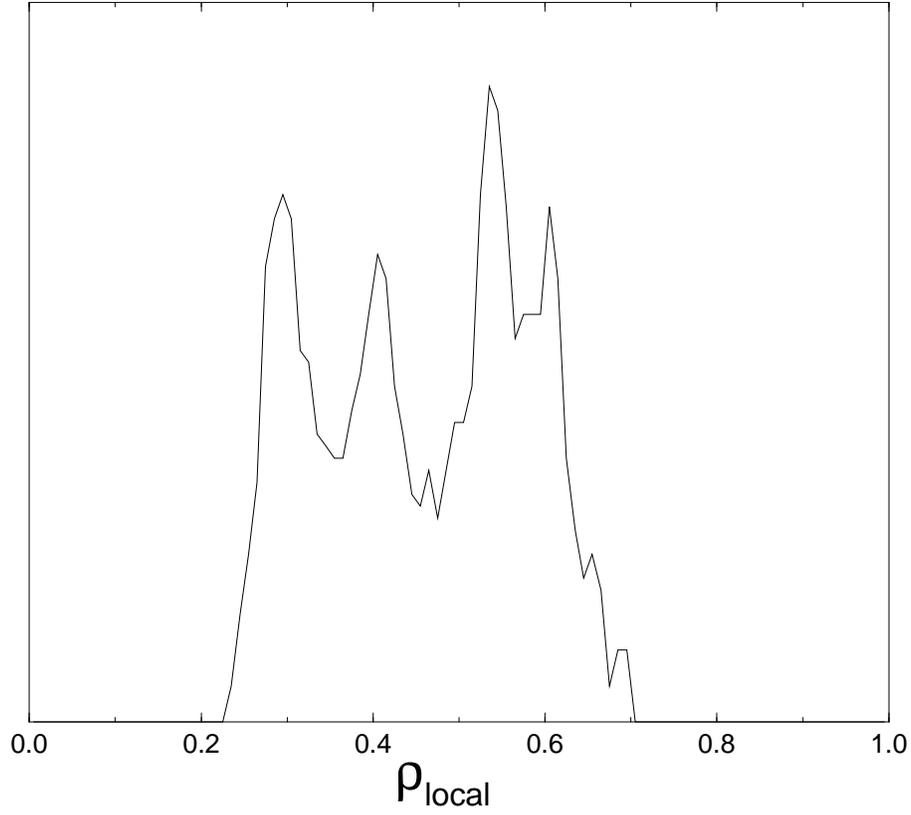,height=5in,angle=-90} 
\vskip 1cm
\caption{Histogram of the local density $\rho_{\rm{local}}$ for $L=1024$ and
  $\rho = 0.46$, averaged over $20^2$ lattice sites.}
\label{peaks}
\end{figure}

The stripe structure breaks down when the system size becomes so small that the
width of a stripe is of the same order as the roughness of its boundary. The
resulting state shows patches of different size, but differs from the critical
state in its cluster size distribution.  In Fig.~\ref{finsize}, the cluster
size distribution for a system at $\rho = 0.46$ in the 'patched' state
($L=512, 1024$), and 'striped' state ($L=2048, 4096$) is plotted. The
transition between the two seems continuous, due to finite--size effects. With
increasing $L$, the bump becomes more distinct because the system tries to
separate one infinite cluster from the ensemble of all clusters. There is no
scaling of $n(s)$, not even in the 'patched' states. With increasing system
size, the bump should separate completely from the size distribution of finite
clusters and move towards $s=\infty$.  The distribution of densities in the
system changes also continuously with increasing $L$, when the transition from
patches to stripes is made.  Fig.~\ref{peaks} shows a histogram of the local
density for $L=1024$, i.e. for the patched state, averaged over $20^2$ sites.
This histogram shows pronounced peaks that are precursors of the stripes.  In
the limit of infinite system size, these peaks will become infinitely sharp.

The densities of the different stripes are related by several equations.  Let
$\rho_1, \ldots, \rho_n$ be the densities in a state with n stripes, starting
with the highest density.  Additionally, we define the density $\rho_{n+1}$ of
the stripe which contained the infinite cluster, immediately after the infinite
cluster has been removed from the system, and before the removed sites are
refilled into the system.  As consequence of a large redistribution, the
different stripes just exchange their densities, i.e.
\begin{equation}
  \rho_{i-1} = \rho_i + (\rho_1 - \rho_{n + 1}) \cdot (1 -
  \rho_i)/(n\,(1 - \rho) + \rho_1 - \rho_{n + 1})
\label{monique}
\end{equation}
for $i = 2, \ldots, n + 1$.  The last factor on the right hand side represents
the fraction of occupied sites of the infinite cluster that are refilled in the
stripe with density $\rho_i$.  We finally obtain
\begin{equation}
\label{ratios}
\frac{1 - \rho_1}{1 - \rho_2} = \frac{1 - \rho_2}{1 - \rho_3} = \ldots
= \frac{1 - \rho_n}{1 - \rho_{n+1}}.
\end{equation}
Together with $\rho = (1/n) \sum_{i=1}^n \rho_i$ we have $n$ equations for
$n+1$ densities.  The average density in a system with $n$ stripes is then
\begin{eqnarray}
  \rho &=& 1 - \frac{1 - \rho_{n+1}}{n} \sum_{i=1}^{n} {\left(\frac{1
      - \rho_1}{1 - \rho_{n+1}}\right)}^{i/n} \nonumber \\ &=& 1 -
  \frac{\rho_1 - \rho_{n+1}}{n(((1-\rho_1)/
    (1-\rho_{n+1}))^{-1/n}-1)}.
\label{noncon}
\end{eqnarray}
This equation was already used in Sec.\ref{criticalregion}.

The striped phase has several features in common with the critical and
subcritical phase: They can all can be characterized by regions of different
density. The density of these regions goes through a cycle, and the spanning
cluster of a region is removed when its density is of the order $\rho^*\approx
62.5\%$.  Although there is no strict synchronization in the subcritical and
critical phase, the system seems to maintain many features of the synchronized
phase, albeit with more irregularities and fluctuations. If there were no
regions with $\rho_{\rm{local}} = \rho^*$, they would be generated by the
same mechanism as in the synchronized phase. The resemblance to the
synchronized phase is especially evident in the one-dimensional case. There,
the critical point can be interpreted as a synchronized state with $\rho^* =
1$, $\rho^\infty = 0$, and an infinite number of stripes.

From our understanding of the striped phase and its stability, we cannot rule
out the existence of gaps, i.e. values of the density for which no synchronized
phase is stable. The following scenario could occur: When decreasing the
density of an $n$--stripe state below the stability threshold, the new density
in stripe 2 after the restructuring could be too large (i.e. $> p_c$) for the
state to be stable.  In this case, the new ($n+1$)-stripe state would be stable
only after lowering the overall density a bit further until $\rho_2 < p_c$. The
intermediate state would then have some irregular patched structure.  However,
in our simulations we could not observe this phenomenon.

\subsection{Hysteresis and Maximum Number of Stripes}
\label{hysteresis}

Since the transition between states with different number of stripes is
discontinuous, hysteresis effects are to be expected.  When the overall density
$\rho$ is decreased adiabatically, a state becomes unstable when $\rho_1$ falls
below $\rho^* \simeq 0.625$, and its number of stripes increases by one
(thereby also increasing $\rho_1$).  When $\rho$ is increased adiabatically,
however, there is no reason, why the system should necessarily rearrange itself
at $\rho_1 = \rho^*$.  The stripe with the highest density can never become
unstable, since it only comes closer to the ideal case of $P(\rho_1) = \rho_1$,
where it consists only of the infinite cluster.  The reorganization from an
$(n+1)$--stripe state to an $n$--stripe state is then triggered by the stripe
with second highest density $\rho_2$.  In the limit of infinite system size, it
takes place when this stripe starts to contain an infinite cluster, i.e. when
$\rho_2$ approaches the percolation threshold. As long as $\rho_2 < p_c$, even
the largest cluster in stripe 2 is small compared to the infinite extent of one
stripe and cannot have any effect on the dynamics.  For finite system size,
however, the reorganization takes place as soon as the ratio $\xi_2 / D$
exceeds a certain critical value, where $D$ is a measure for the linear extent
of one synchronized region (as e.g. the thickness of a stripe in the
two-dimensional system with periodic boundary conditions).  Therefore, the
effect of hysteresis in a finite system is difficult to observe when the number
of stripes $n$ increases.  Only for the transition from the two-stripe state to
the homogeneous state we could identify a significant interval of hysteresis.
The two-stripe state could be kept alive up to $\rho = 0.64$ for $L = 4096$.

In the following, we argue that there exists an upper limit to the number of
stripes $n$ even in an infinitely large system.  When lowering $\rho$, the
difference $\rho_1 - \rho_2$ decreases with each additional stripe. If the
number of stripes $n$ was not bounded from above, we would have $\rho_1 -
\rho_2 \to 0$ and $\rho_1 \to \rho^*$.  Such a state could only be stable if
the maximum cluster size was finite in all stripes but stripe 1, which, in
turn, could only be possible if the percolation threshold was identical to
$\rho^*$.  Since the distribution of occupied sites within a stripe is not
completely random, the percolation threshold can be different from the
threshold for conventional site percolation. We can find its value by
simulating the density sequence $(\rho_n, \rho_{n-1},...,\rho_1)$ for a single
stripe.  We start with a stationary state with density $\rho$ just above
$\rho^*$. We remove the infinite cluster and fill the system again randomly and
continually, and we measure the percentage of the largest cluster for two
different system sizes. The results are shown in Tab.~\ref{thresholdchange}.
We see from Tab.~\ref{thresholdchange} that an infinite cluster exists already
for $\rho = 0.60$, since $P$ is non-vanishing and does not decrease with
increasing $L$.  Furthermore, we measured the correlation length of the finite
clusters in the interval $0.57 \le \rho \le 0.62$, which shows a pronounced
peak around $\rho = 0.59$ for both $L = 512$ and $L = 1024$ (see
Tab.~\ref{thresholdchange}).  These results show that the percolation threshold
for stripes is very close (if not identical) to $p_c$ for site percolation.
Thus, states with $\rho_2 > 0.59$ cannot exist, and there must exist a minimum
density gap $\Delta \rho = \rho_1 - \rho_2$, and, consequently, a finite
maximum number $N$ of stripes.  Eq.~(\ref{ratios}) gives a maximum possible
number of $n = 11(\pm 2)$ stripes, and a corresponding minimum mean density
$\rho = 42(\pm 0.3)\%$. In our simulations, due to finite system size, we could
observe a maximum number of $n=5$ stripes at $\rho = 0.45$ and $L = 4096$ (see
Fig.~\ref{fivephases}). For other lattice types, the maximum number of stripes
and the densities where the phase transitions take place are of course
different.  The realization of an infinite number of stripes (i.e. a front
moving through a continuum) is possible with different rules, where the
occupied sites of the chosen cluster are removed one by one and are put back
into the system, before the next site is removed (see \cite{cla96}).

\begin{figure}
\psfig{file=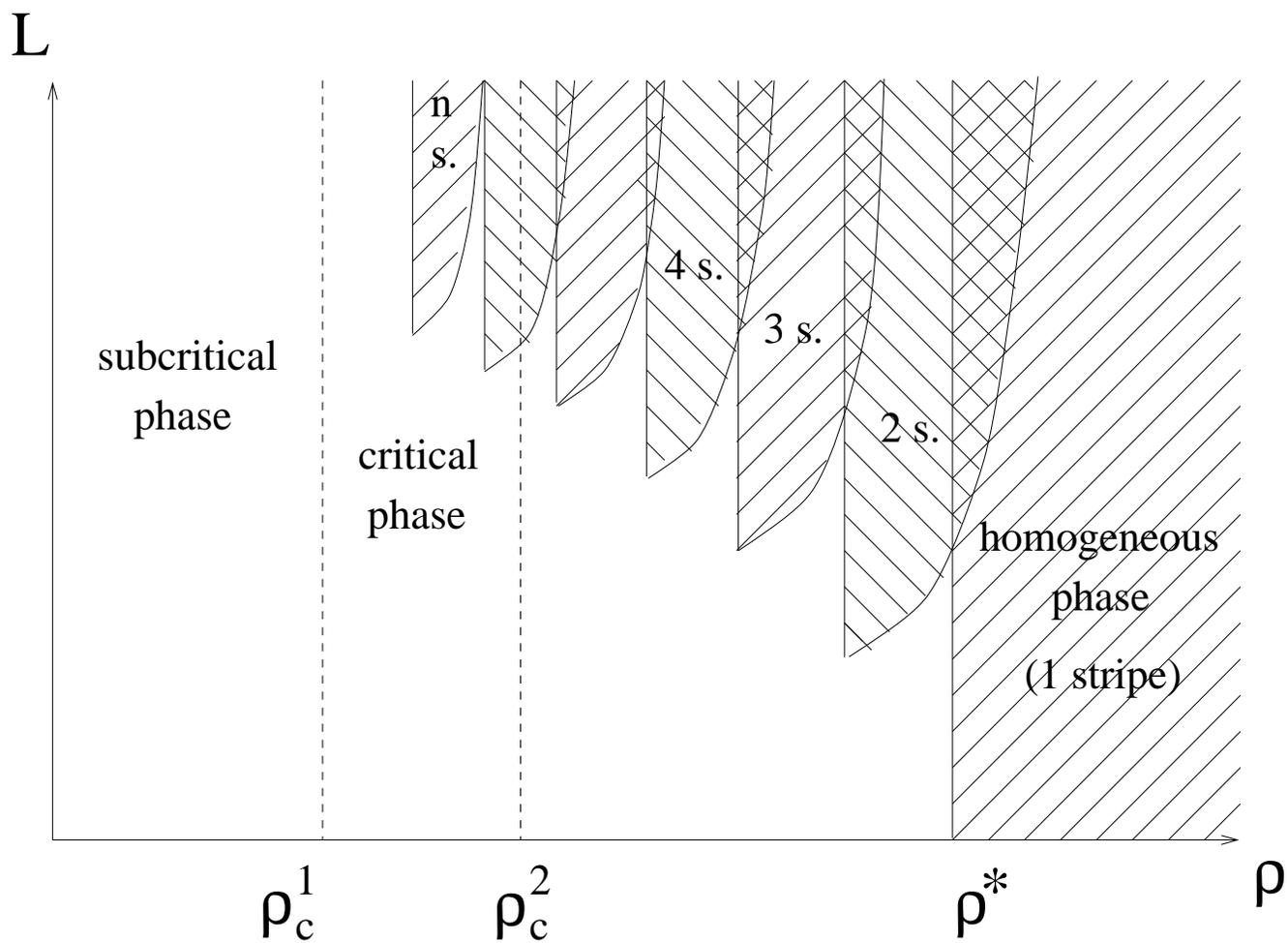,height=5in,angle=-90} 
\vskip 1cm
\caption{Schematic phase diagram of a two--dimensional square system for all
  densities from 0 to 1}
\label{phasediagram}
\end{figure}

A schematic phase diagram of the system for all densities from 0 to 1 is shown
in Fig.~\ref{phasediagram}.  Since a state with $n$ stripes is only stable if 
the thickness of a stripe is much larger than $\xi_2$, the phase boundaries 
depend on the system size $L$ for small $L$. They become vertical for large
$L$, and the values of $\rho$, where the phase
transitions take place, are well defined.  As one can also see in
Fig.~\ref{phasediagram}, the minimum density of the synchronized phase is
smaller than the upper limit $\rho_c^{(2)}$ of the critical phase.  Since the
transition from patches to stripes is discontinuous, it shows also hysteresis.

\subsection{Stability, Movement, and Roughness of Stripes}
\label{roughness}

In systems with $n > 3$ stripes, there exists the possibility that the stripes
might not be arranged in consecutive order with respect to their densities.
However, these arrangements are not stable, as will be shown in the following.
If the stripes are arranged in consecutive order, the stripe with highest
density has always the same densities in its neighboring stripes during the
complete cycle of length $n$.  Each time the infinite cluster is removed, the
borders of the stripe it has occupied expand $\xi_2$ to one side and $\xi_n$ to
the other.  Since during one cycle the infinite cluster occupies each stripe
position exactly once, all changes of width cancel. Since $\xi_2 > \xi_n$, the
whole pattern moves into the direction of the stripe with second highest
density, as seen from the infinite cluster. If the stripes are not arranged in
consecutive order, e.g. $\{ \rho_1, \rho_3, \rho_2, \rho_4 \}$, the changes in
width do not cancel. The left neighbors of the infinite cluster during one
cycle are $\rho_4, \rho_3, \rho_3, \rho_2$, and its right neighbors $\rho_3,
\rho_2, \rho_4, \rho_3$. Therefore, there is a net change of the width of some
stripes on the expense of others, and sooner or later (depending on system
size) the structure breaks down, and the system rearranges itself to a state
with the stripes arranged in consecutive order. The instability of certain
configurations like the one mentioned above was tested by starting with an
artificially generated stationary state with the 'wrong' order of stripes.
After some time, the restructuring to the stable stationary state could be
observed.

Similarly, if the area of different stripes is different or if the values of
the stripe densities are not the same after each rearrangement of occupied
sites, different stripes see a different environment when they are removed.
Such states can therefore no be stable. However, when they are close enough to
the stable state, they return to it, as follows from the observed stability of
striped states.  In order to check explicitly the stability with respect to
variations in the densities in a system with two stripes, we start with an
unperturbed state at density $\rho^0$ with two stripes of density $\rho_1^0$
and $\rho_2^0$, respectively.  The strength of the infinite cluster is $P^0 =
P(\rho_1^0)$, and $\rho_1^0 - P^0 = \rho_3^0$.  Now we perturb the system by
changing the densities to $\rho_1 = \rho_1^0 + \delta$ and $\rho_2 = \rho_2^0 -
\delta$ with small $\delta > 0$.  The overall density $\rho^0$ is kept
constant.  The strength of the infinite cluster is now $P = P(\rho_1) =
P(\rho_1^0 + \delta) = P^0 + \delta \cdot P'(\rho_1^0)$. After the removal of
the infinite cluster the densities are $\rho_3 = \rho_1 - P = \rho_3^0 - \delta
\cdot (P' - 1)$ and $\rho_2 = \rho_2^0 - \delta$.  Using Eq.~(\ref{monique})
one obtains for the highest density in the next iteration step
\[
  \rho_1^{\rm{new}} = \rho_2^0 - \delta + (P^0 + \delta \cdot P')
  \frac{1 - \rho_2^0 + \delta}{2 - (\rho_2^0 + \rho_3^0) + \delta
    \cdot P'}.
\]
If this density is lower than the perturbed density $\rho_1 = \rho_1^0 +
\delta$, the system wants to return to the unperturbed configuration and the
initial state is stable. For stability, thus, the following inequality has to
hold:
\[
  \rho_2^0 - \delta + (P^0 + \delta \cdot P') \frac{1 - \rho_2^0 +
    \delta}{2 - (\rho_2^0 + \rho_3^0) + \delta \cdot P'} < \rho_2^0 +
  P^0 \frac{1 - \rho_2^0}{2 - (\rho_2^0 + \rho_3^0)} + \delta.
\]
On the right hand side, we have used $\rho_1^0 = \rho_2^0 + P^0 (1 -
\rho_2^0)/(2 - (\rho_2^0 + \rho_3^0))$ from Eq.~(\ref{monique}).  This
inequality can be rewritten as
\begin{equation}
  P' < \frac{4 (1 - \rho^0) + 3 P^0}{(1 - P^0/(2 (1 - \rho^0 + P^0)))
    (1 - \rho_2^0)}
\label{sharyl}
\end{equation}
We measured all quantities that appear in Eq.~(\ref{sharyl}) in the density
interval where the two-stripe state could be observed and found that it is
completely contained in the interval where this condition is fulfilled.

\begin{figure}
\vskip 1cm
\psfig{file=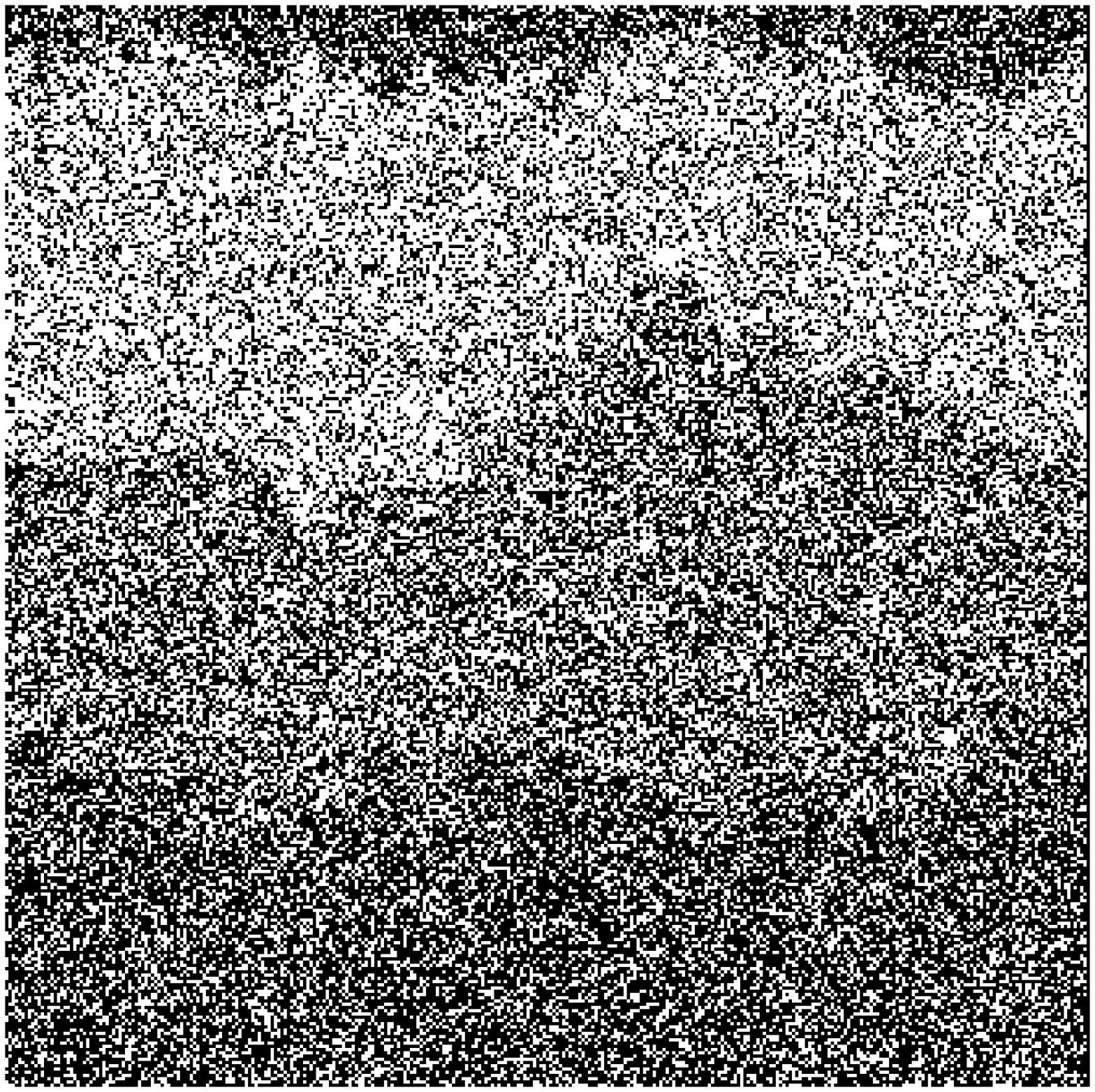,width=2.65cm}
\vskip -2.7cm
\hskip 2.75cm
\psfig{file=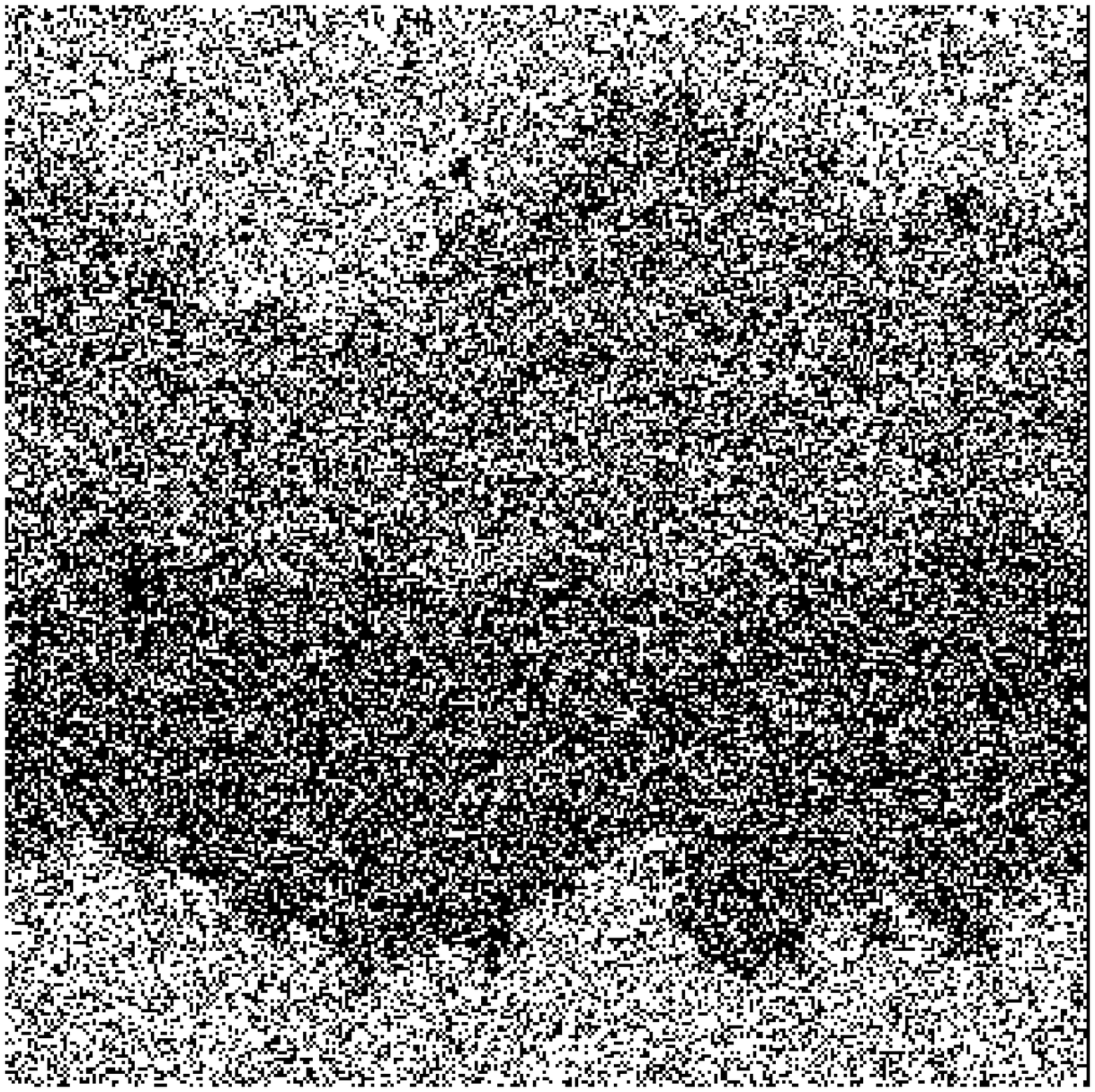,width=2.65cm}
\hskip 0.5cm
\psfig{file=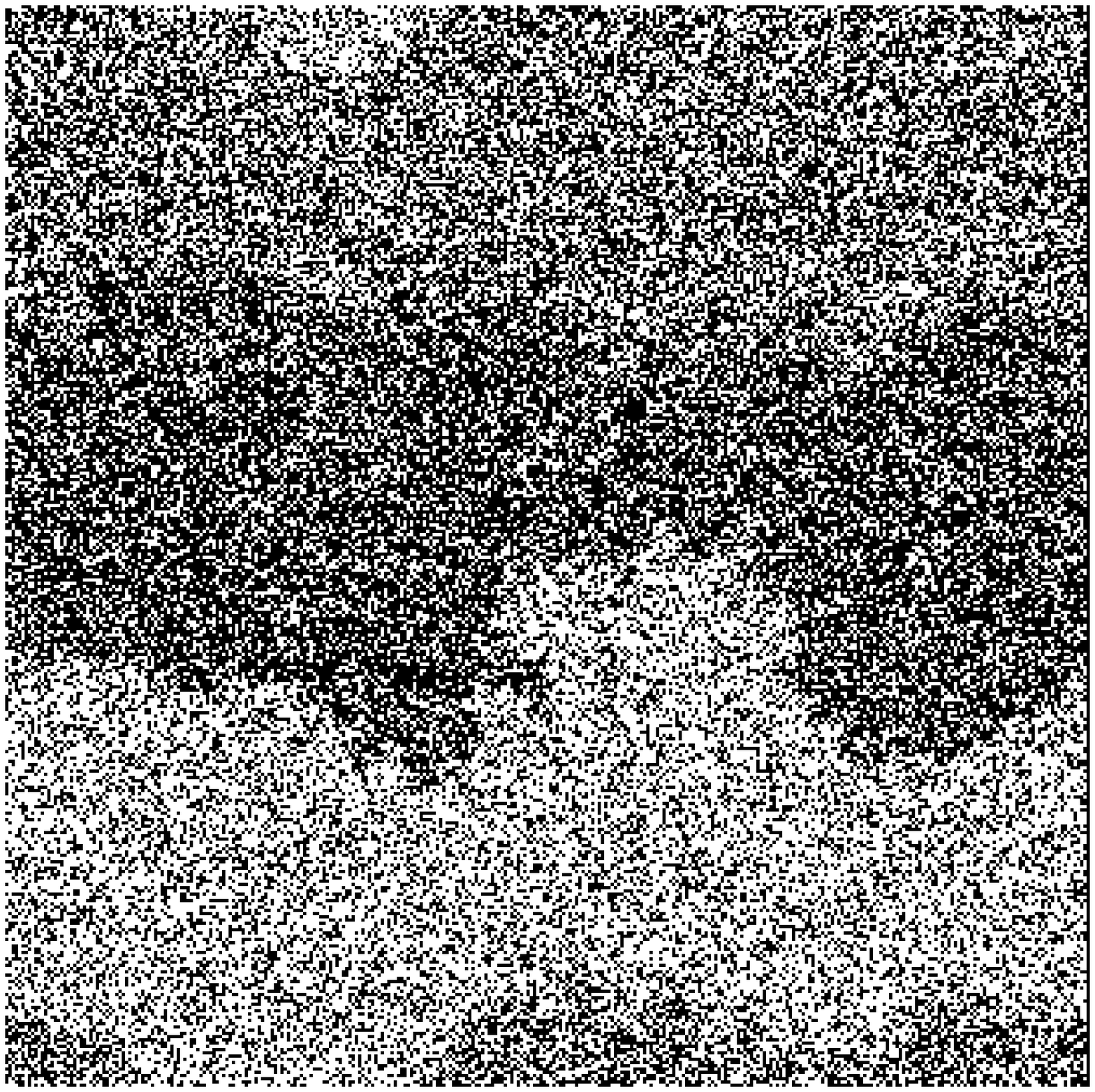,width=2.65cm}
\hskip 0.5cm
\psfig{file=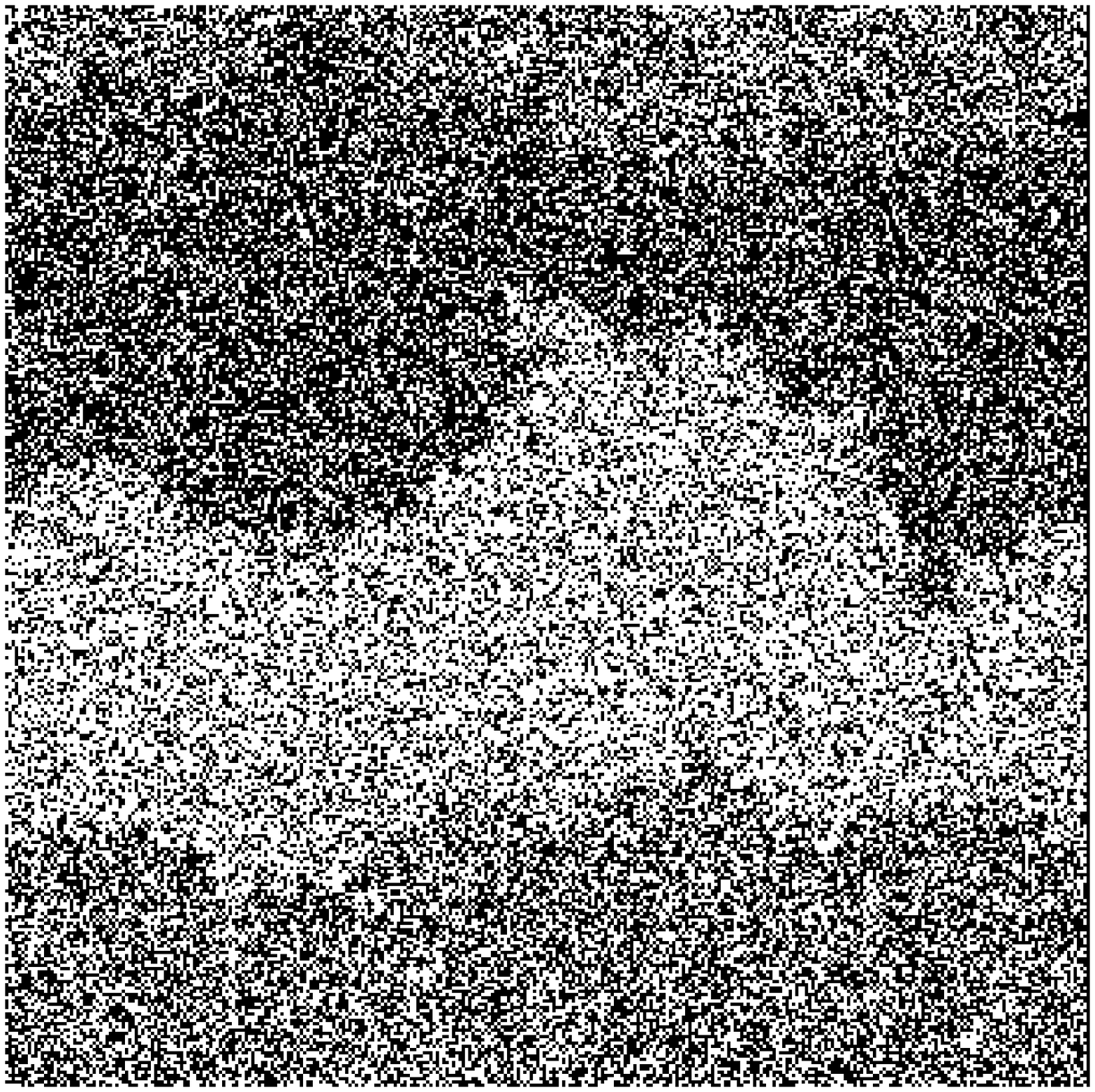,width=2.65cm}
\hskip 0.5cm
\psfig{file=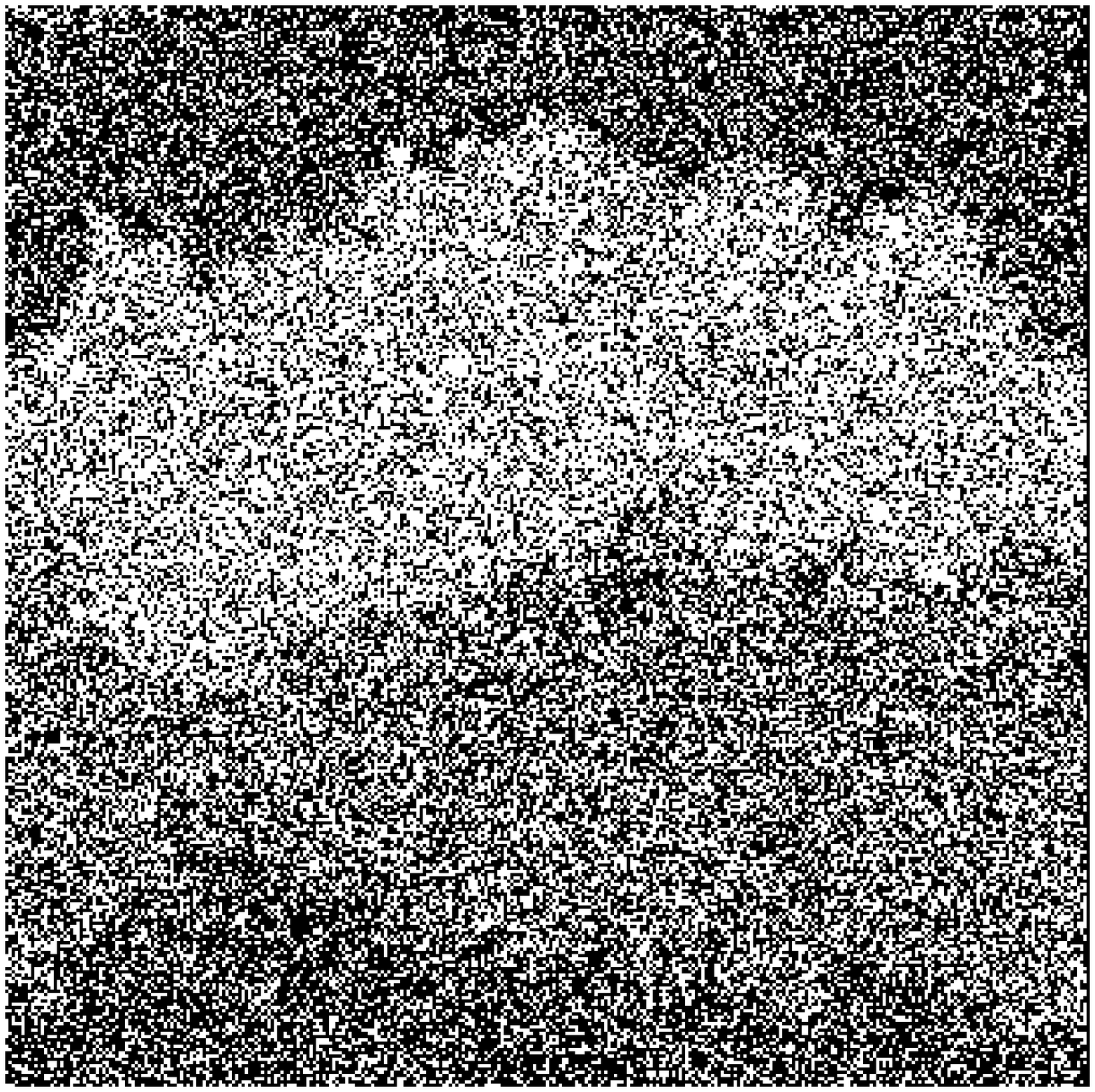,width=2.65cm}
\vskip 1cm
\caption{A sequence of stationary states of a three-stripe state at $\rho =
  50\%$ and $L = 1024$. The snapshots were taken at times $t_0$, $t_0+51$,
  $t_0+102$, $t_0+153$ and $t_0+204$.}
\label{veloc}
\end{figure}

As already mentioned above, the whole pattern of stripes moves into the
direction of the stripe with second highest density seen from the infinite
cluster. This movement can be observed in the simulations for all states with
$n > 2$. The net velocity in a two-stripe state is zero due to symmetry.  We
measured the velocity of this movement for stationary states with three and
four stripes at $\rho = 50\%$ and $46\%$. The results are $v \approx 4.3 \pm
0.5$ lattice sites per iteration step for $\rho = 50\%$ and $v \approx 5.2 \pm
0.5$ for $\rho = 46\%$. Since the degree to which the infinite cluster reaches
into the neighboring stripes depends on their correlation lengths, the velocity
is essentially determined by the density in stripe 2. As one can see from
Tab.~\ref{tabstripe}, $\rho_2$ is larger for $\rho = 46\%$ than for $\rho =
50\%$, therefore the velocity is higher for $\rho = 46\%$.  The movement of the
stripe boundaries can be seen in Fig.~\ref{veloc}, where a sequence of
stationary states of a three-stripe state is shown.

As one can see in Fig.~\ref{fivephases}, the stripes seem to have rough rather
than smooth boundaries.  When the infinite cluster is removed, all finite
clusters in the adjacent stripes that are connected to it are also removed.
This leads to not only to a net velocity in one preferred direction, but also
to a roughening of the stripe boundaries, even in situations where the
(artificially generated) initial state consists of strictly horizontal stripes.

The roughness of the interface is characterized by an exponent $\alpha$ that
describes how the saturation width of the interface $w_{\rm{sat}}$ scales
with its length $L$,
\[
  w_{\rm{sat}}(L) \propto L^\alpha.
\]
The interface width $w$ is defined as the root mean square of the differences
$y(x) - \bar y(x)$ where $y(x)$ ($\bar y(x)$) is the (averaged) $y$--coordinate
of the interface (assuming horizontal stripes)
\[
  w(L, t) = \sqrt{\frac{1}{L} \sum_{i=1}^L [y(x,t) - \bar y(t)]^2},
\]
with $\bar y(t) = (1/L) \sum_{i=1}^L y(x,t)$.  We assume that $y(t)$ is
single-valued, i.e. there are no overhangs.  This assumption is always correct
at sufficiently large scales.

To characterize the time-dependent dynamics of the roughening of the interface
width $w(t)$, the growth exponent $\beta$ is introduced by
\[
  w(L, t) \propto t^\beta.
\]
For small $t$, there is no dependence on $L$, since the information on system
size needs a certain time to spread.  For an introduction to interface
roughening, see \cite{bar95}.

We measured the roughness exponent $\alpha$ as well as the growth exponent
$\beta$ for stationary states with two and three stripes at $\rho = 55\%$ and
$50\%$. States with a larger number of stripes can only survive in large
systems, so we were not able to scan a broad enough range of system sizes $L$
to determine an exponent. Since at sufficiently large scales there is an
up-down symmetry in the case $n=2$, which is absent in the case $n=3$, where
the boundary moves in one preferred direction, we expect different exponents.
 
\begin{figure}
\psfig{file=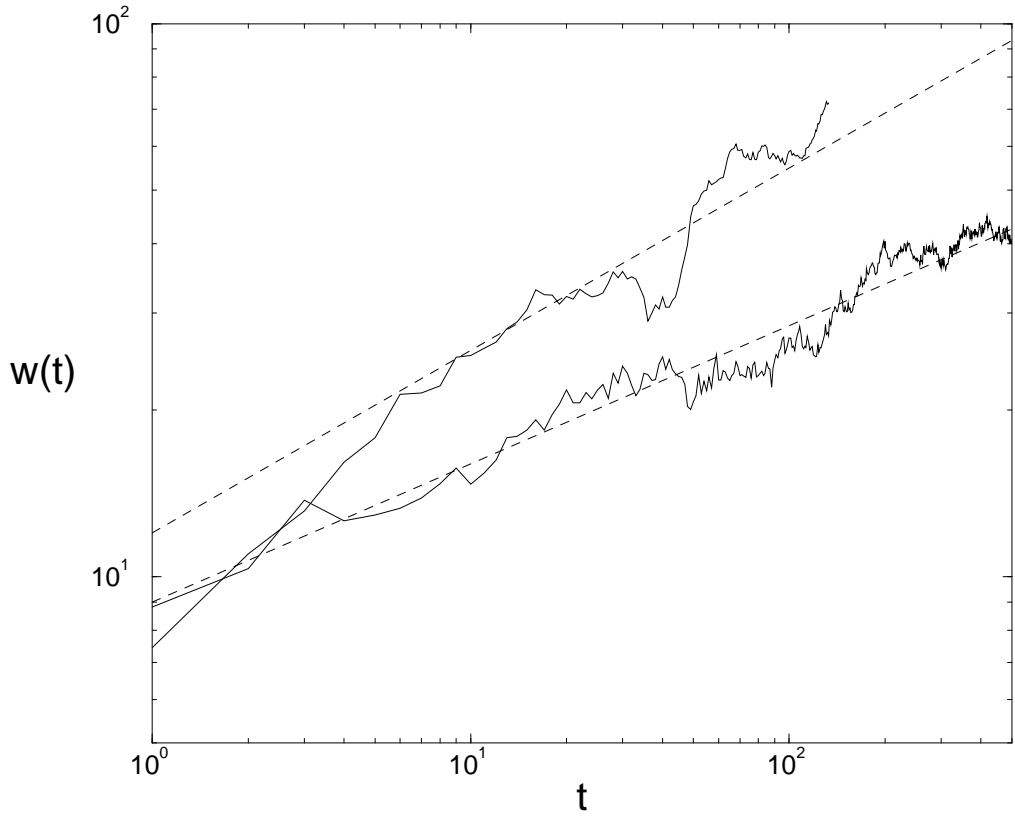,height=5in,angle=-90} 
\vskip 1cm
\caption{Width $w$ of the stripe boundary in the cases $n=2$ (lower curve) and
  $n=3$ (upper curve) for small $t$. The dashed lines have slope $1/4$ and
  $1/3$, respectively.}
\label{beta}
\end{figure}

The results for $\alpha$ and $\beta$ are shown in Tab.~\ref{tabstripe}.
Fig.~\ref{beta} shows the roughening of the boundary for the cases $n=2$ and
$n=3$. The simulation results are compatible with the values $\alpha =
1/2, \beta = 1/3$ for $n=3$ and $\alpha = 1/2, \beta = 1/4$ for $n=2$, 
although other values cannot be ruled out. The
former exponents are those of the Kardar-Parisi-Zhang universality class
\cite{kar86}, whereas the latter belong to the Edwards-Wilkinson universality
class \cite{edw82}, describing an interface in thermal equilibrium.  For all $n
\ge 3$, the boundaries exhibit a net movement in one preferred direction, thus
breaking the up-down symmetry. Therefore we expect to find the KPZ universality
class also in the cases $n\ge 4$ that could not be investigated in the
simulations.

\begin{figure}
\psfig{file=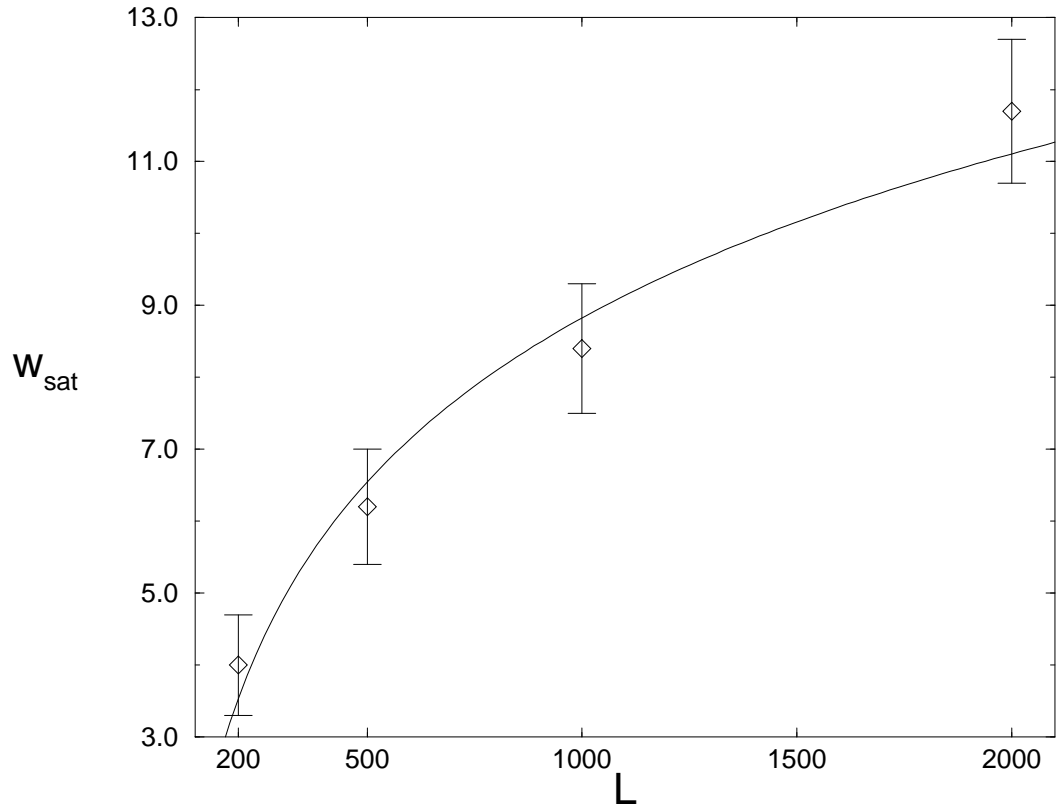,height=5in,angle=-90} 
\vskip 1cm
\caption{Saturation width $w_{\rm{sat}}$ of an excitation front at $\rho =
  45\%$ and $\rho^{\rm{before}} = 74\%$ ($L = 200$, 500, 1000, 2000). The
  smooth curve is a logarithmic fit.}
\label{alpha}
\end{figure}

In the limit of a large number of stripes $n\to\infty$, the movement of the
infinite cluster is reminiscent of the movement of a front.  To study the
dynamics of such a front, we introduce a third state ('excited') that marks the
sites belonging to the front.  The excitation spreads at each time step to all
of its nearest occupied neighbors leaving behind an empty site. In order to
assure a constant density, one has to re-occupy randomly chosen empty sites.
Starting with a flat excitation front in an initial state of average density
$\rho$, after short time the front will either have disappeared or the system
will have evolved to a stationary state, where the front propagates
quasi-deterministically in one direction. After leaving the system at one end,
it reenters at the opposite end.
Although the front state can be interpreted as a stripe state with $n=\infty$,
there is one important difference. In the stripe state, the infinite cluster
sees a medium with density below $p_c$ in front of it, whereas in the front
state, this density is above $p_c$. One can therefore not expect to find KPZ
behavior as in the stripe state. Instead, the shortest paths (using only
occupied sites) from the sites of the front at time $t = 0$ to the sites of the
front at some later time have a Gaussian distribution, leading to a width of
the interface that scales as $\log(L)$, i.e., $\alpha = 0$. Fig.~\ref{alpha}
shows the saturation width as function of the system size, together with a
logarithmic fit.

One can also compare this behavior to the behavior of an excitation front in a
percolation system without periodic boundary conditions perpendicular to the
front, i.e. a system of size $L$ parallel to the front and of size '$\infty$'
perpendicular to it. The 'infinite' extension perpendicular to the front
would in practice be realized by providing a constant, homogeneous density
ahead of the front, such that the front will never pass through a region where
it already has been before, and additional correlations due to sites left over
from the last passing are eliminated.  In this case, the stability arguments of
Subsec.~\ref{synchro} and of \cite{cla96}, which lead to the instability for
$\rho^{\rm{before}} < \rho^*$, are no longer valid, and one can investigate
the properties of the front in the whole density interval of
$\rho^{\rm{before}} \in [p_c; 1]$. Then, one can also measure the exponent
that describes how the velocity of the front vanishes when the density
$\rho^{\rm{before}}$ approaches $p_c$.  As in the previous paragraph, we
expect that such a front has a roughness exponent $\alpha=0$. Models which
yield a front of finite roughness must therefore be more complicated and
include a dependence of the front propagation on the shape of the front, or
memory effects.  Such models, as the Kuramoto--Shivashinsky equation, or the
model in \cite{pro96}, give a KPZ roughness exponent.

\section{Summary and Conclusions}
\label{conclusion}

To conclude, we have described a nonequilibrium percolation model which shows
several new phenomena which are unknown in equilibrium percolation. Clusters of
occupied sites are removed and refilled into the system at randomly chosen
empty sites.  For densities smaller than a critical density, there are only
finite clusters, as in percolation theory. At the critical point, the critical
exponents assume values different from percolation theory, and they do not
satisfy a hyperscaling relation. For densities between the critical density
$\approx 40.8\%$ and a second density $\approx 43.5\%$, the system remains
critical, with the correlation length diverging slower than the system size.
The value of the exponent that relates the correlation length with the system
size depends on the density. For densities larger than $\approx 43.5\%$, the
system has a finite number of regions of different densities. The number of
regions depends on the density, and the transitions between states with
different numbers of regions are discontinuous and show hysteresis. The shape
of the regions is striped for periodic boundary conditions.

We were able to describe the dynamics of the system in the critical state as
well as in the striped state in terms of patches of different densities. On
length scales smaller than a 'finger thickness' (critical region) or the
stripe thickness (synchronized phase), the density of a patch is fairly
homogeneous and goes through a temporal cycle: Starting from a small density,
the density increases until most occupied sites of the patch are removed at
$\rho \approx 62.5\%$, and the cycle restarts. The system is therefore
synchronized on small length scales. In the critical interval, the density is
not large enough to allow a synchronization over the total width of the system,
leading to the observed power laws. A similar relation between critical
behavior and incomplete synchronization was found in the earthquake model
\cite{mid95}.  There, complete synchronization is hindered by the boundary
conditions. In our model, the global density conservation prevents
synchronization over distances larger than some correlation length. In both
models, the correlation length diverges slower than the system size.  A model
with similar phenomena (i.e., subcritical, critical, and supercritical phases)
has been reported in \cite{cor97}.

The density cycles observed in our model cannot exist in equilibrium systems,
which are invariant under time reversal. Therefore, fluctuations in an
equilibrium system are usually small and well described by a Gauss distribution
around some mean value (except for the neighborhood of a critical point where
the next--order terms have to be added to the free--energy functional). One
consequence is the extensivity of equilibrium systems, i.e. a part of an
equilibrium system behaves like a smaller system of the size of the part. This
is not true for our model.

\begin{figure}
\psfig{file=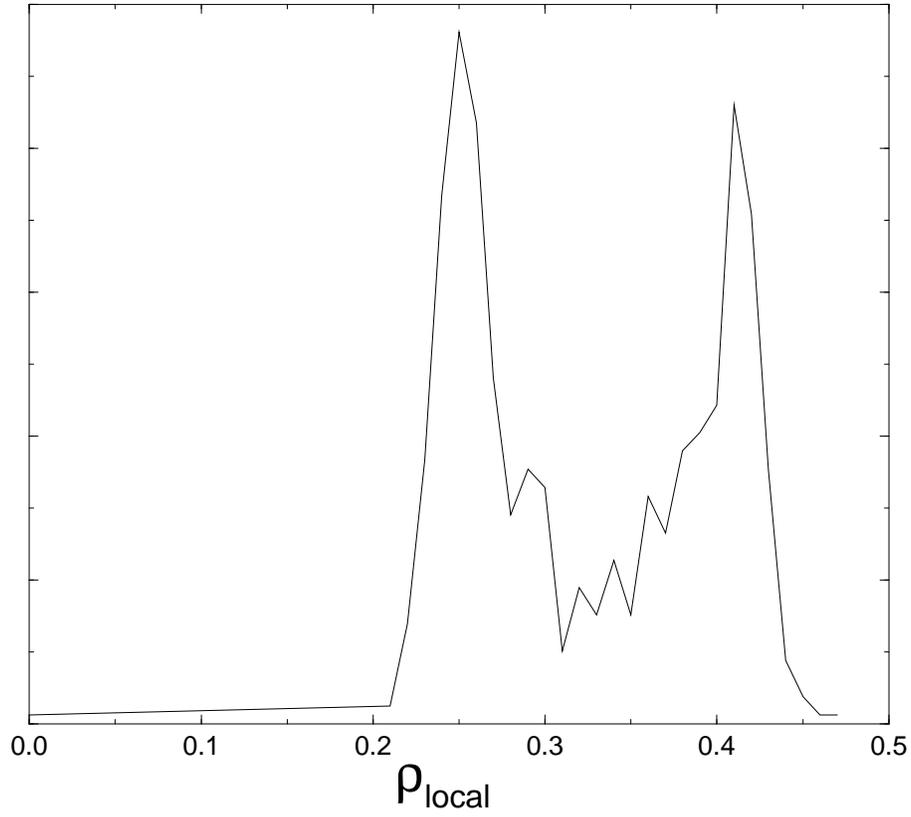,height=5in,angle=-90}
\vskip 1cm
\caption{Histogram of the local density $\rho_{\rm{local}}$ in a
  three-dimensional system for $L=100$ and $\rho \approx 0.34$ in arbitrary
  units, averaged over $11^3$ lattice sites.}
\label{threedee}
\end{figure}

Although our simulations were performed in two dimensions, we expect the same
general picture in higher dimensions: Unfortunately, in dimensions $d > 2$, one
has to cope with severe finite-size effects.  For example, the
three-dimensional analogue of a two-stripe system would be a two-layer system.
For this state to be stable, the correlation lengths in the layers have to be
small compared to their thickness. With a given maximum number of sites
$\approx 1.6 \times 10^7$ that could be simulated, the maximum thickness of one
of two layers in a three-dimensional system is only $\approx 125$ compared to
$\approx 2000$ in two dimensions. Another problem is that a visualization of
the states like in two dimensions is not possible.  One has therefore to
measure local densities in order to determine the number of stripes that
coexist in a given state. For the three-dimensional system, we could verify the
transition to two stripes at $\rho^*_{\rm{3D}} \approx 0.34$.  In
Fig.~\ref{threedee}, we plotted a histogram of the local density in a system at
$\rho \lesssim \rho^*_{\rm{3D}}$, proving the existence of two stripes.

We also simulated a variant where not an arbitrary site is selected, but always
a site that belongs to the largest cluster in the system.  The simulations were
carried out for system sizes up to $L = 1024$.  Although the snapshots look
different, since small clusters are no longer removed, we did not find
different behavior.  The important quantities, i.e., the critical exponents,
$\rho_c^{(1)}$, $\rho_c^{(2)}$ and $\rho^*$ remain the same.

An interesting modification of the model does not conserve the density of
occupied sites exactly, but only on the average. While such a modification
would not change the behavior of an equilibrium system, we expect some
important changes in our model. When the density is not strictly conserved,
large global density fluctuations can occur, and a sharp distinction between a
subcritical, a critical, and a supercritical phase is no longer possible.  This
new model will be treated in another publication \cite{cla97}.

The existence of these fundamental differences between equilibrium and many
nonequilibrium systems makes it difficult to apply methods developed for the
study of equilibrium critical behavior in nonequilibrium systems. It remains
still a challenge to find a field--theoretical formalism that allows to analyze
analytically the critical behavior of this and related models.

S.C. was supported by the Deutsche Forschungsgemeinschaft (DFG)
Contract No Schw 348/7-1.
B.D. was supported by EPSRC Grant GR/K79307.

\newpage

\begin{table}
\begin{tabular}{llllll}
  $\rho$ & 0.41 & 0.42 & 0.43 & 0.435 \\ \tableline $\phi_1$ & 1.46(8)
  & 1.72(6) & 1.86(8) & 1.92(8) \\ $\phi_2$ & 0.79(2) & 0.92(3) &
  0.97(3) & 0.99(2) \\ $\phi_3$ & 1.23(3) & 1.50(3) & 1.62(3) &
  1.69(5)
\end{tabular}
\caption{The exponents $\phi_{1,2,3}$ of $s_{\rm{max}}$, $\xi$, and 
  $S$ for various densities $\rho_c^{(1)}<\rho<\rho_c^{(2)}$ ($L =
  512$, 1024, 2048, 4096)}
\label{fsexpo}
\end{table}

\begin{table}
\begin{tabular}{lrrrrrr}
  $\rho$ & 0.57 & 0.58 & 0.59 & 0.60 & 0.61 & 0.62 \\ \tableline $P(L
  = 512)$ & - & - & - & 0.33 & 0.46 & 0.52 \\ $P(L = 1024)$ & - & - &
  - & 0.34 & 0.47 & 0.53 \\ $\xi(L = 512)$ & 57 & 85 & 133 & 62 & 36 &
  13 \\ $\xi(L = 1024)$ & 58 & 89 & 145 & 75 & 28 & 14
\end{tabular}
\caption{The percentage $P$ of the largest cluster and the correlation
  length $\xi$ for different densities $\rho = 0.57, \ldots, 0.62$
  with $L = 512$ and $L = 1024$}
\label{thresholdchange}
\end{table}

\begin{table}
\begin{tabular}{llll}
  $\rho$ & 0.46 & 0.50 & 0.55 \\ \tableline Nr. of stripes & 4 & 3 & 2
  \\ $\rho^1$ & 0.63 & 0.66 & 0.66 \\ $\rho^2$ & 0.53 & 0.51 & 0.44 \\ 
  $\alpha$ & - & 0.44(10) & 0.46(10) \\ $\beta$ & - & 0.35(5) & 0.25(5) \\ 
  $v$ & 5.2(5) & 4.3(5) & 0
\end{tabular}
\caption{Roughness exponent $\alpha$, growth exponent $\beta$, and velocity
  $v$ of the stripes for various densities ($200 \le L \le 2000$).}
\label{tabstripe}
\end{table}

\end{document}

%% file: psfig_19.tex
\def\PsfigVersion{1.9}
\ifx\undefined\psfig\else \fi

%

\let\LaTeXAtSign=\@
\let\@=\relax
\edef\psfigRestoreAt{\catcode`\@=\number\catcode`@\relax}
\catcode`\@=11\relax
\newwrite\@unused
\def\ps@typeout#1{{\let\protect\string\immediate\write\@unused{#1}}}
\ps@typeout{psfig/tex \PsfigVersion}


\def\figurepath{./}
\def\psfigurepath#1{\edef\figurepath{#1}}

%
%
\def\@nnil{\@nil}
\def\@empty{}
\def\@psdonoop#1\@@#2#3{}
\def\@psdo#1:=#2\do#3{\edef\@psdotmp{#2}\ifx\@psdotmp\@empty \else
    \expandafter\@psdoloop#2,\@nil,\@nil\@@#1{#3}\fi}
\def\@psdoloop#1,#2,#3\@@#4#5{\def#4{#1}\ifx #4\@nnil \else
       #5\def#4{#2}\ifx #4\@nnil \else#5\@ipsdoloop #3\@@#4{#5}\fi\fi}
\def\@ipsdoloop#1,#2\@@#3#4{\def#3{#1}\ifx #3\@nnil 
       \let\@nextwhile=\@psdonoop \else
      #4\relax\let\@nextwhile=\@ipsdoloop\fi\@nextwhile#2\@@#3{#4}}
\def\@tpsdo#1:=#2\do#3{\xdef\@psdotmp{#2}\ifx\@psdotmp\@empty \else
    \@tpsdoloop#2\@nil\@nil\@@#1{#3}\fi}
\def\@tpsdoloop#1#2\@@#3#4{\def#3{#1}\ifx #3\@nnil 
       \let\@nextwhile=\@psdonoop \else
      #4\relax\let\@nextwhile=\@tpsdoloop\fi\@nextwhile#2\@@#3{#4}}
%
\ifx\undefined\fbox
\newdimen\fboxrule
\newdimen\fboxsep
\newdimen\ps@tempdima
\newbox\ps@tempboxa
\fboxsep = 3pt
\fboxrule = .4pt
\long\def\fbox#1{\leavevmode\setbox\ps@tempboxa\hbox{#1}\ps@tempdima\fboxrule
    \advance\ps@tempdima \fboxsep \advance\ps@tempdima \dp\ps@tempboxa
   \hbox{\lower \ps@tempdima\hbox
  {\vbox{\hrule height \fboxrule
          \hbox{\vrule width \fboxrule \hskip\fboxsep
          \vbox{\vskip\fboxsep \box\ps@tempboxa\vskip\fboxsep}\hskip 
                 \fboxsep\vrule width \fboxrule}
                 \hrule height \fboxrule}}}}
\fi
%
%
\newread\ps@stream
\newif\ifnot@eof       
\newif\if@noisy        
\newif\if@atend        
\newif\if@psfile       
%
%
{\catcode`\%=12\global\gdef\epsf@start{
\def\epsf@PS{PS}
\def\epsf@getbb#1{%
%
%
\openin\ps@stream=#1
\ifeof\ps@stream\ps@typeout{Error, File #1 not found}\else
%
%
   {\not@eoftrue \chardef\other=12
    \def\do##1{\catcode`##1=\other}\dospecials \catcode`\ =10
    \loop
       \if@psfile
	  \read\ps@stream to \epsf@fileline
       \else{
	  \obeyspaces
          \read\ps@stream to \epsf@tmp\global\let\epsf@fileline\epsf@tmp}
       \fi
       \ifeof\ps@stream\not@eoffalse\else
%
%
       \if@psfile\else
       \expandafter\epsf@test\epsf@fileline:. \\%
       \fi
%
%
          \expandafter\epsf@aux\epsf@fileline:. \\%
       \fi
   \ifnot@eof\repeat
   }\closein\ps@stream\fi}%
%
%
\long\def\epsf@test#1#2#3:#4\\{\def\epsf@testit{#1#2}
			\ifx\epsf@testit\epsf@start\else
\ps@typeout{Warning! File does not start with `\epsf@start'.  It may not be a PostScript file.}
			\fi
			\@psfiletrue} 
%
%
{\catcode`\%=12\global\let\epsf@percent=
%
%
%
\long\def\epsf@aux#1#2:#3\\{\ifx#1\epsf@percent
   \def\epsf@testit{#2}\ifx\epsf@testit\epsf@bblit
	\@atendfalse
        \epsf@atend #3 . \\%
	\if@atend	
	   \if@verbose{
		\ps@typeout{psfig: found `(atend)'; continuing search}
	   }\fi
        \else
        \epsf@grab #3 . . . \\%
        \not@eoffalse
        \global\no@bbfalse
        \fi
   \fi\fi}%
%
%
\def\epsf@grab #1 #2 #3 #4 #5\\{%
   \global\def\epsf@llx{#1}\ifx\epsf@llx\empty
      \epsf@grab #2 #3 #4 #5 .\\\else
   \global\def\epsf@lly{#2}%
   \global\def\epsf@urx{#3}\global\def\epsf@ury{#4}\fi}%
%
%
\def\epsf@atendlit{(atend)} 
\def\epsf@atend #1 #2 #3\\{%
   \def\epsf@tmp{#1}\ifx\epsf@tmp\empty
      \epsf@atend #2 #3 .\\\else
   \ifx\epsf@tmp\epsf@atendlit\@atendtrue\fi\fi}


\chardef\psletter = 11 
\chardef\other = 12

\newif \ifdebug 
\newif\ifc@mpute 
\c@mputetrue 

\let\then = \relax
\def\r@dian{pt }
\let\r@dians = \r@dian
\let\dimensionless@nit = \r@dian
\let\dimensionless@nits = \dimensionless@nit
\def\internal@nit{sp }
\let\internal@nits = \internal@nit
\newif\ifstillc@nverging
\def \Mess@ge #1{\ifdebug \then \message {#1} \fi}

{ 
	\catcode `\@ = \psletter
	\gdef \nodimen {\expandafter \n@dimen \the \dimen}
	\gdef \term #1 #2 #3%
	       {\edef \t@ {\the #1}
		\edef \t@@ {\expandafter \n@dimen \the #2\r@dian}%
		\t@rm {\t@} {\t@@} {#3}%
	       }
	\gdef \t@rm #1 #2 #3%
	       {{%
		\count 0 = 0
		\dimen 0 = 1 \dimensionless@nit
		\dimen 2 = #2\relax
		\Mess@ge {Calculating term #1 of \nodimen 2}%
		\loop
		\ifnum	\count 0 < #1
		\then	\advance \count 0 by 1
			\Mess@ge {Iteration \the \count 0 \space}%
			\Multiply \dimen 0 by {\dimen 2}%
			\Mess@ge {After multiplication, term = \nodimen 0}%
			\Divide \dimen 0 by {\count 0}%
			\Mess@ge {After division, term = \nodimen 0}%
		\repeat
		\Mess@ge {Final value for term #1 of 
				\nodimen 2 \space is \nodimen 0}%
		\xdef \Term {#3 = \nodimen 0 \r@dians}%
		\aftergroup \Term
	       }}
	\catcode `\p = \other
	\catcode `\t = \other
	\gdef \n@dimen #1pt{#1} 
}

\def \Divide #1by #2{\divide #1 by #2} 

\def \Multiply #1by #2
       {{
	\count 0 = #1\relax
	\count 2 = #2\relax
	\count 4 = 65536
	\Mess@ge {Before scaling, count 0 = \the \count 0 \space and
			count 2 = \the \count 2}%
	\ifnum	\count 0 > 32767 
	\then	\divide \count 0 by 4
		\divide \count 4 by 4
	\else	\ifnum	\count 0 < -32767
		\then	\divide \count 0 by 4
			\divide \count 4 by 4
		\else
		\fi
	\fi
	\ifnum	\count 2 > 32767 
	\then	\divide \count 2 by 4
		\divide \count 4 by 4
	\else	\ifnum	\count 2 < -32767
		\then	\divide \count 2 by 4
			\divide \count 4 by 4
		\else
		\fi
	\fi
	\multiply \count 0 by \count 2
	\divide \count 0 by \count 4
	\xdef \product {#1 = \the \count 0 \internal@nits}%
	\aftergroup \product
       }}

\def\r@duce{\ifdim\dimen0 > 90\r@dian \then   
		\multiply\dimen0 by -1
		\advance\dimen0 by 180\r@dian
		\r@duce
	    \else \ifdim\dimen0 < -90\r@dian \then  
		\advance\dimen0 by 360\r@dian
		\r@duce
		\fi
	    \fi}

\def\Sine#1%
       {{%
	\dimen 0 = #1 \r@dian
	\r@duce
	\ifdim\dimen0 = -90\r@dian \then
	   \dimen4 = -1\r@dian
	   \c@mputefalse
	\fi
	\ifdim\dimen0 = 90\r@dian \then
	   \dimen4 = 1\r@dian
	   \c@mputefalse
	\fi
	\ifdim\dimen0 = 0\r@dian \then
	   \dimen4 = 0\r@dian
	   \c@mputefalse
	\fi
	\ifc@mpute \then
		\divide\dimen0 by 180
		\dimen0=3.141592654\dimen0
		\dimen 2 = 3.1415926535897963\r@dian 
		\divide\dimen 2 by 2 
		\Mess@ge {Sin: calculating Sin of \nodimen 0}%
		\count 0 = 1 
		\dimen 2 = 1 \r@dian 
		\dimen 4 = 0 \r@dian 
		\loop
			\ifnum	\dimen 2 = 0 
			\then	\stillc@nvergingfalse 
			\else	\stillc@nvergingtrue
			\fi
			\ifstillc@nverging 
			\then	\term {\count 0} {\dimen 0} {\dimen 2}%
				\advance \count 0 by 2
				\count 2 = \count 0
				\divide \count 2 by 2
				\ifodd	\count 2 
				\then	\advance \dimen 4 by \dimen 2
				\else	\advance \dimen 4 by -\dimen 2
				\fi
		\repeat
	\fi		
			\xdef \sine {\nodimen 4}%
       }}

\def\Cosine#1{\ifx\sine\UnDefined\edef\Savesine{\relax}\else
		             \edef\Savesine{\sine}\fi
	{\dimen0=#1\r@dian\advance\dimen0 by 90\r@dian
	 \Sine{\nodimen 0}
	 \xdef\cosine{\sine}
	 \xdef\sine{\Savesine}}}	      

\def\psdraft{
	\def\@psdraft{0}
}
\def\psfull{
	\def\@psdraft{100}
}

\psfull

\newif\if@scalefirst
\def\psscalefirst{\@scalefirsttrue}
\def\psrotatefirst{\@scalefirstfalse}
\psrotatefirst

\newif\if@draftbox
\def\psnodraftbox{
	\@draftboxfalse
}
\def\psdraftbox{
	\@draftboxtrue
}
\@draftboxtrue

\newif\if@prologfile
\newif\if@postlogfile
\def\pssilent{
	\@noisyfalse
}
\def\psnoisy{
	\@noisytrue
}
\psnoisy
\newif\if@bbllx
\newif\if@bblly
\newif\if@bburx
\newif\if@bbury
\newif\if@height
\newif\if@width
\newif\if@rheight
\newif\if@rwidth
\newif\if@angle
\newif\if@clip
\newif\if@verbose
\def\@p@@sclip#1{\@cliptrue}

\newif\if@decmpr


\def\@p@@sfigure#1{\def\@p@sfile{null}\def\@p@sbbfile{null}
	        \openin1=#1.bb
		\ifeof1\closein1
	        	\openin1=\figurepath#1.bb
			\ifeof1\closein1
			        \openin1=#1
				\ifeof1\closein1%
				       \openin1=\figurepath#1
					\ifeof1
					   \ps@typeout{Error, File #1 not found}
						\if@bbllx\if@bblly
				   		\if@bburx\if@bbury
			      				\def\@p@sfile{#1}%
			      				\def\@p@sbbfile{#1}%
							\@decmprfalse
				  	   	\fi\fi\fi\fi
					\else\closein1
				    		\def\@p@sfile{\figurepath#1}%
				    		\def\@p@sbbfile{\figurepath#1}%
						\@decmprfalse
	                       		\fi%
			 	\else\closein1%
					\def\@p@sfile{#1}
					\def\@p@sbbfile{#1}
					\@decmprfalse
			 	\fi
			\else
				\def\@p@sfile{\figurepath#1}
				\def\@p@sbbfile{\figurepath#1.bb}
				\@decmprtrue
			\fi
		\else
			\def\@p@sfile{#1}
			\def\@p@sbbfile{#1.bb}
			\@decmprtrue
		\fi}

\def\@p@@sfile#1{\@p@@sfigure{#1}}

\def\@p@@sbbllx#1{
		\@bbllxtrue
		\dimen100=#1
		\edef\@p@sbbllx{\number\dimen100}
}
\def\@p@@sbblly#1{
		\@bbllytrue
		\dimen100=#1
		\edef\@p@sbblly{\number\dimen100}
}
\def\@p@@sbburx#1{
		\@bburxtrue
		\dimen100=#1
		\edef\@p@sbburx{\number\dimen100}
}
\def\@p@@sbbury#1{
		\@bburytrue
		\dimen100=#1
		\edef\@p@sbbury{\number\dimen100}
}
\def\@p@@sheight#1{
		\@heighttrue
		\dimen100=#1
   		\edef\@p@sheight{\number\dimen100}
}
\def\@p@@swidth#1{
		\@widthtrue
		\dimen100=#1
		\edef\@p@swidth{\number\dimen100}
}
\def\@p@@srheight#1{
		\@rheighttrue
		\dimen100=#1
		\edef\@p@srheight{\number\dimen100}
}
\def\@p@@srwidth#1{
		\@rwidthtrue
		\dimen100=#1
		\edef\@p@srwidth{\number\dimen100}
}
\def\@p@@sangle#1{
		\@angletrue
		\edef\@p@sangle{#1} 
}
\def\@p@@ssilent#1{ 
		\@verbosefalse
}
\def\@p@@sprolog#1{\@prologfiletrue\def\@prologfileval{#1}}
\def\@p@@spostlog#1{\@postlogfiletrue\def\@postlogfileval{#1}}
\def\@cs@name#1{\csname #1\endcsname}
\def\@setparms#1=#2,{\@cs@name{@p@@s#1}{#2}}
%
%
\def\ps@init@parms{
		\@bbllxfalse \@bbllyfalse
		\@bburxfalse \@bburyfalse
		\@heightfalse \@widthfalse
		\@rheightfalse \@rwidthfalse
		\def\@p@sbbllx{}\def\@p@sbblly{}
		\def\@p@sbburx{}\def\@p@sbbury{}
		\def\@p@sheight{}\def\@p@swidth{}
		\def\@p@srheight{}\def\@p@srwidth{}
		\def\@p@sangle{0}
		\def\@p@sfile{} \def\@p@sbbfile{}
		\def\@p@scost{10}
		\def\@sc{}
		\@prologfilefalse
		\@postlogfilefalse
		\@clipfalse
		\if@noisy
			\@verbosetrue
		\else
			\@verbosefalse
		\fi
}
%
%
\def\parse@ps@parms#1{
	 	\@psdo\@psfiga:=#1\do
		   {\expandafter\@setparms\@psfiga,}}
%
%
\newif\ifno@bb
\def\bb@missing{
	\if@verbose{
		\ps@typeout{psfig: searching \@p@sbbfile \space  for bounding box}
	}\fi
	\no@bbtrue
	\epsf@getbb{\@p@sbbfile}
        \ifno@bb \else \bb@cull\epsf@llx\epsf@lly\epsf@urx\epsf@ury\fi
}	
\def\bb@cull#1#2#3#4{
	\dimen100=#1 bp\edef\@p@sbbllx{\number\dimen100}
	\dimen100=#2 bp\edef\@p@sbblly{\number\dimen100}
	\dimen100=#3 bp\edef\@p@sbburx{\number\dimen100}
	\dimen100=#4 bp\edef\@p@sbbury{\number\dimen100}
	\no@bbfalse
}
\newdimen\p@intvaluex
\newdimen\p@intvaluey
\def\rotate@#1#2{{\dimen0=#1 sp\dimen1=#2 sp
		  \global\p@intvaluex=\cosine\dimen0
		  \dimen3=\sine\dimen1
		  \global\advance\p@intvaluex by -\dimen3
		  \global\p@intvaluey=\sine\dimen0
		  \dimen3=\cosine\dimen1
		  \global\advance\p@intvaluey by \dimen3
		  }}
\def\compute@bb{
		\no@bbfalse
		\if@bbllx \else \no@bbtrue \fi
		\if@bblly \else \no@bbtrue \fi
		\if@bburx \else \no@bbtrue \fi
		\if@bbury \else \no@bbtrue \fi
		\ifno@bb \bb@missing \fi
		\ifno@bb \ps@typeout{FATAL ERROR: no bb supplied or found}
			\no-bb-error
		\fi
		%
%
		\count203=\@p@sbburx
		\count204=\@p@sbbury
		\advance\count203 by -\@p@sbbllx
		\advance\count204 by -\@p@sbblly
		\edef\ps@bbw{\number\count203}
		\edef\ps@bbh{\number\count204}
		\if@angle 
			\Sine{\@p@sangle}\Cosine{\@p@sangle}
	        	{\dimen100=\maxdimen\xdef\r@p@sbbllx{\number\dimen100}
					    \xdef\r@p@sbblly{\number\dimen100}
			                    \xdef\r@p@sbburx{-\number\dimen100}
					    \xdef\r@p@sbbury{-\number\dimen100}}
%
                        \def\minmaxtest{
			   \ifnum\number\p@intvaluex<\r@p@sbbllx
			      \xdef\r@p@sbbllx{\number\p@intvaluex}\fi
			   \ifnum\number\p@intvaluex>\r@p@sbburx
			      \xdef\r@p@sbburx{\number\p@intvaluex}\fi
			   \ifnum\number\p@intvaluey<\r@p@sbblly
			      \xdef\r@p@sbblly{\number\p@intvaluey}\fi
			   \ifnum\number\p@intvaluey>\r@p@sbbury
			      \xdef\r@p@sbbury{\number\p@intvaluey}\fi
			   }
			\rotate@{\@p@sbbllx}{\@p@sbblly}
			\minmaxtest
			\rotate@{\@p@sbbllx}{\@p@sbbury}
			\minmaxtest
			\rotate@{\@p@sbburx}{\@p@sbblly}
			\minmaxtest
			\rotate@{\@p@sbburx}{\@p@sbbury}
			\minmaxtest
			\edef\@p@sbbllx{\r@p@sbbllx}\edef\@p@sbblly{\r@p@sbblly}
			\edef\@p@sbburx{\r@p@sbburx}\edef\@p@sbbury{\r@p@sbbury}
		\fi
		\count203=\@p@sbburx
		\count204=\@p@sbbury
		\advance\count203 by -\@p@sbbllx
		\advance\count204 by -\@p@sbblly
		\edef\@bbw{\number\count203}
		\edef\@bbh{\number\count204}
}
%
%
\def\in@hundreds#1#2#3{\count240=#2 \count241=#3
		     \count100=\count240	
		     \divide\count100 by \count241
		     \count101=\count100
		     \multiply\count101 by \count241
		     \advance\count240 by -\count101
		     \multiply\count240 by 10
		     \count101=\count240	
		     \divide\count101 by \count241
		     \count102=\count101
		     \multiply\count102 by \count241
		     \advance\count240 by -\count102
		     \multiply\count240 by 10
		     \count102=\count240	
		     \divide\count102 by \count241
		     \count200=#1\count205=0
		     \count201=\count200
			\multiply\count201 by \count100
		 	\advance\count205 by \count201
		     \count201=\count200
			\divide\count201 by 10
			\multiply\count201 by \count101
			\advance\count205 by \count201
		     \count201=\count200
			\divide\count201 by 100
			\multiply\count201 by \count102
			\advance\count205 by \count201
		     \edef\@result{\number\count205}
}
\def\compute@wfromh{
		\in@hundreds{\@p@sheight}{\@bbw}{\@bbh}
		\edef\@p@swidth{\@result}
}
\def\compute@hfromw{
	        \in@hundreds{\@p@swidth}{\@bbh}{\@bbw}
		\edef\@p@sheight{\@result}
}
\def\compute@handw{
		\if@height 
			\if@width
			\else
				\compute@wfromh
			\fi
		\else 
			\if@width
				\compute@hfromw
			\else
				\edef\@p@sheight{\@bbh}
				\edef\@p@swidth{\@bbw}
			\fi
		\fi
}
\def\compute@resv{
		\if@rheight \else \edef\@p@srheight{\@p@sheight} \fi
		\if@rwidth \else \edef\@p@srwidth{\@p@swidth} \fi
}
%
\def\compute@sizes{
	\compute@bb
	\if@scalefirst\if@angle
	\if@width
	   \in@hundreds{\@p@swidth}{\@bbw}{\ps@bbw}
	   \edef\@p@swidth{\@result}
	\fi
	\if@height
	   \in@hundreds{\@p@sheight}{\@bbh}{\ps@bbh}
	   \edef\@p@sheight{\@result}
	\fi
	\fi\fi
	\compute@handw
	\compute@resv}

%
%
\def\psfig#1{\vbox {
	%
	\ps@init@parms
	\parse@ps@parms{#1}
	\compute@sizes
	\ifnum\@p@scost<\@psdraft{
		\special{ps::[begin] 	\@p@swidth \space \@p@sheight \space
				\@p@sbbllx \space \@p@sbblly \space
				\@p@sbburx \space \@p@sbbury \space
				startTexFig \space }
		\if@angle
			\special {ps:: \@p@sangle \space rotate \space} 
		\fi
		\if@clip{
			\if@verbose{
				\ps@typeout{(clip)}
			}\fi
			\special{ps:: doclip \space }
		}\fi
		\if@prologfile
		    \special{ps: plotfile \@prologfileval \space } \fi
		\if@decmpr{
			\if@verbose{
				\ps@typeout{psfig: including \@p@sfile.Z \space }
			}\fi
			\special{ps: plotfile "`zcat \@p@sfile.Z" \space }
		}\else{
			\if@verbose{
				\ps@typeout{psfig: including \@p@sfile \space }
			}\fi
			\special{ps: plotfile \@p@sfile \space }
		}\fi
		\if@postlogfile
		    \special{ps: plotfile \@postlogfileval \space } \fi
		\special{ps::[end] endTexFig \space }
		\vbox to \@p@srheight sp{
			\hbox to \@p@srwidth sp{
				\hss
			}
		\vss
		}
	}\else{
		\if@draftbox{		
			\hbox{\frame{\vbox to \@p@srheight sp{
			\vss
			\hbox to \@p@srwidth sp{ \hss \@p@sfile \hss }
			\vss
			}}}
		}\else{
			\vbox to \@p@srheight sp{
			\vss
			\hbox to \@p@srwidth sp{\hss}
			\vss
			}
		}\fi

	}\fi
}}
\psfigRestoreAt
\let\@=\LaTeXAtSign